\documentclass{article}

\usepackage[final]{neurips_2026}

\usepackage[normalem]{ulem}
\usepackage{graphicx}
\usepackage{subcaption}
\usepackage{url}
\usepackage[breaklinks=true,colorlinks=true,linkcolor=black,citecolor=black,urlcolor=blue]{hyperref}
\usepackage{fancyhdr}

\usepackage{amsthm}
\usepackage{amsmath}
\usepackage{multirow}
\usepackage{tcolorbox}
\usepackage{algorithm}
\usepackage{algorithmicx}
\usepackage{algpseudocode}
\usepackage{enumitem}
\usepackage{makecell}
\usepackage{booktabs}
\usepackage{pifont}
\usepackage{colortbl}

\usepackage{amssymb}
\usepackage{bbm}

\newcommand{\SystemName}{\textsc{Spotlight}}
\newcommand{\Bes}{\textsc{veRL-omni(spot)}}

\newcommand{\Bs}{\textsc{RLBoost}}
\newcommand{\Be}{\textsc{veRL-omni(3x)}}
\newcommand{\B}{\textsc{RLBoost(3x)}}

\newcommand{\posttrain}{post-training}
\newcommand{\tcache}{Teacache}
\newcommand{\ocr}{DeepSeek-OCR}
\newcommand{\geneval}{Geneval}

\newcommand{\PHM}[1]{\vspace{.4em}\noindent\textbf{#1}}

\usepackage{tikz}

\theoremstyle{definition}

\ifdefined\RELEASE
  \newcommand{\ignore}[1]{}
  \newcommand{\fixme}[1]{}
  \newcommand{\dmi}[1]{}
  \newcommand{\rui}[1]{}
  \newcommand{\dak}[1]{}
  \newcommand{\leo}[1]{}
  \newcommand{\kev}[1]{}
  \newcommand{\myai}[1]{}
  \newcommand{\rod}[1]{}
  \newcommand{\TODO}[1]{}
\else
  \newcommand{\ignore}[1]{}
  \newcommand{\fixme}[1]{{\textcolor{red}{[~FIXME:~#1~]}}}
  \newcommand{\dmi}[1]{{\textcolor{blue}{[~D:~#1~]}}}
  \newcommand{\rui}[1]{{\textcolor{orange}{[~R:~#1~]}}}
  \newcommand{\dak}[1]{{\textcolor{magenta}{[~DAK:~#1~]}}}
  
  \newcommand{\leo}[1]{{\textcolor{teal}{[~L:~#1~]}}}
  \newcommand{\kev}[1]{{\textcolor{brown}{[~K:~#1~]}}}
  
  \newcommand{\myai}[1]{{\textcolor{olive}{[~AI:~#1~]}}}
  \newcommand{\rod}[1]{{\textcolor{green}{[~R:~#1~]}}}
  \newcommand{\TODO}[1]{{\textcolor{red}{TODO:~#1}}}
\fi

\title{\SystemName{}: Synergizing Seed Exploration and Spot GPUs for DiT RL Post-Training}

\author{
  \textbf{Ruiqi Lai}$^{1*}$,
  \textbf{Dakai An}$^{2*}$,
  \textbf{Wei Gao}$^{2}$,
  \textbf{Ju Huang}$^{3}$,
  \textbf{Siran Yang}$^{3}$,\\
  \textbf{Jiamang Wang}$^{3}$,
  \textbf{Lin Qu}$^{3}$,
  \textbf{Dmitrii Ustiugov}$^{1\dagger}$,
  \textbf{Wei Wang}$^{2}$ \\[0.3em]
  $^1$NTU Singapore \\
  $^2$Hong Kong University of Science and Technology \\
  $^3$Alibaba Group \\[0.3em]
  \texttt{RUIQI003@e.ntu.edu.sg, \{danab, csgaowei, weiwa\}@cse.ust.hk,} \\
  \texttt{\{huangju.hj, siran.ysr, jiamang.wang\}@alibaba-inc.com,} \\
  \texttt{xide.ql@taobao.com, dmitrii.ustiugov@ntu.edu.sg}
}

\begin{document}

\makeatletter
\renewcommand{\@noticestring}{}
\makeatother

\fancypagestyle{firstpagestyle}{%
  \fancyhf{}%
  \fancyhead[L]{\includegraphics[height=20pt]{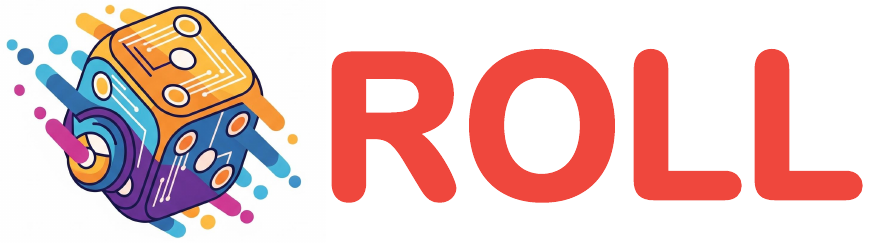}}%
  \renewcommand{\headrulewidth}{0pt}%
  \setlength{\headheight}{40pt}%
  \fancyfoot[C]{\thepage}%
}

\maketitle
\thispagestyle{firstpagestyle}

\renewcommand{\thefootnote}{}\footnotetext{$^\ast$Equal contribution. $^\dagger$Corresponding author.}\renewcommand{\thefootnote}{\arabic{footnote}}

\begin{abstract}
Reinforcement learning (RL) post-training of Diffusion Transformers (DiTs) is prohibitively expensive, requiring thousands of high-end GPUs. Existing works explore two directions to reduce cost: seed exploration improves training convergence by selecting high-contrast samples, yet adds compute to the critical path; spot GPUs offer 69--77\% lower cost, yet sit idle during training because DiT rollouts finish nearly simultaneously, which prevents LLM-style pipelining of rollout with training. Spot preemptions further break Sequence Parallelism (SP) groups, fragmenting GPU topology.

We present \SystemName{}, the first system that harvests spot GPUs for DiT RL post-training. \SystemName{} rests on two key insights we devise: (1)~we show that exploration can tolerate stale model weights because
exploration that uses the model weights from the previous iteration preserves the relative ranking of random seeds, allowing exploration to run on idle spot GPUs during training. (2)~SP reconfiguration can reuse on-node state, reducing group recovery from minutes to sub-second launches. Built on these insights, \SystemName{} introduces three techniques: a bandit-based exploration planner that maximizes reward variance within the training time budget, elastic sequence parallelism that reconfigures SP groups on the fly via persistent schedulers and intra-node weight copying, and a preemption-aware pull-based request scheduler that balances load and commits in-flight state upon preemption. We implement \SystemName{} on the open-source RL platform ROLL and evaluate it on Qwen-Image post-training. \SystemName{} reaches the same target validation score $4\times$ faster than baselines, reducing total cost by $1.4$-$6.4\times$ while achieving superior image quality on DeepSeek-OCR and Geneval datasets with resolution $512\times512$ and $1280\times1280$.
\end{abstract}

\section{Introduction}
\begin{figure}[t]
    \centering
    \includegraphics[width=\linewidth]{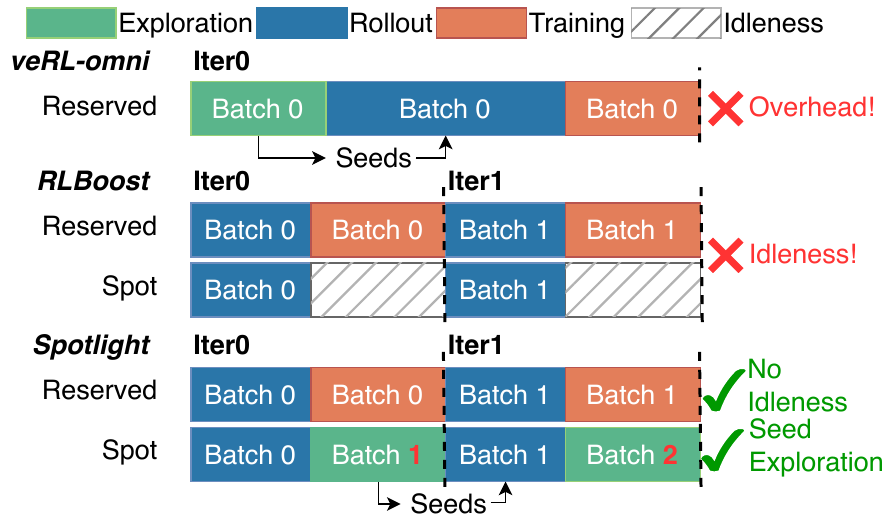}
    \vspace{-20pt}
    \caption{Overview of existing works and \SystemName{}. \SystemName{} breaks the dependency of the exploration phase on current model weights, overlapping exploration and rollout phases.}
    \label{fig:intro}
    \vspace{-1em}
\end{figure}

Diffusion Transformers (DiTs) have become a dominant architecture for generating high-quality images and videos, and have been widely deployed in production systems such as SeeDream~\cite{seedream}, Sora~\cite{sora}, SeedDance~\cite{seeddance}, and HappyHorse~\cite{happyhorse}. Frontier labs~\cite{stepvideo,seedream} further adopt DiT RL post-training to improve generation quality via GRPO-style algorithms~\cite{dancegrpo,flowgrpo}. 
A typical iteration consists of two phases: \emph{rollout} phase, which generates $K$ samples per prompt using distinct random seeds to form a \emph{prompt group}, followed by scoring each sample; 
and \emph{training}, which computes gradients from the generated samples and corresponding scores, then synchronizes model weights back to the rollout phase for the next iteration. Both the rollout and training phases are highly compute-intensive~\cite{tetriserve,golden2024generativeaillmsimplications,ma2025stepvideot2vtechnicalreportpractice}. Thus, these labs dedicate thousands of GPUs to post-training their diffusion models~\cite{stepvideo,cui2025emu35nativemultimodalmodels} for extended periods, rendering the training cost prohibitively high. This necessitates a cost-efficient DiT RL post-training system.

One way to reduce cost is to adopt \emph{seed exploration} to accelerate training convergence and reduce the number of iterations needed to reach the target accuracy~\cite{dancegrpo,treegrpo,branchgrpo,sol-rl}. Since the random seed affects the generation quality of diffusion models, these methods explore seeds within a prompt group to find those with the highest and lowest reward scores, thereby maximizing reward variance and obtaining more informative gradients. However, this exploration phase lies on the critical path of each iteration and incurs additional computation overhead, as shown in the first row of Fig.~\ref{fig:intro}, which increases iteration time. As a result, it may offset the gains from faster convergence and undermine cost-effectiveness.


Another, more cost-effective way is to leverage cheap preemptible resources, such as spot GPUs. Spot GPUs provide access to spare datacenter capacity at 69--77\% lower cost~\cite{rlboost}, but can be reclaimed at short notice~\cite{aws_grace,gcp_grace} and offer limited inter-node bandwidth. Prior LLM RL post-training systems such as RLBoost place rollout on spot GPUs and training on reserved GPUs, since rollout is embarrassingly parallel while training requires stable collective communication. This design, however, does not transfer directly to DiT RL: diffusion rollouts exhibit tightly clustered completion times, leaving almost no stragglers whose partially completed samples could be overlapped with training. The synchronization barrier between rollout and training thus leaves spot GPUs idle for the entire training phase. Beyond this idle time, spot GPUs pose an additional challenge. \emph{Sequence Parallelism} (SP) is widely adopted to run heavy DiT workloads across multiple GPUs, yet on spot pools frequent preemptions and reallocations break SP groups and fragment the GPU topology, wasting a significant portion of resources.

A striking opportunity hides in plain sight: seed exploration is starved for compute on the critical path, while harvesting spot GPUs leaves spot capacity idle during the training phase. Although existing cost-efficient RL systems~\cite{rlboost,rlhfless,dancegrpo,treegrpo,branchgrpo,sol-rl} pursue the two in isolation, they overlook this synergy. We devise two key insights to harness it. 

\PHM{Insight 1. Exploration tolerates stale model weights.} Exploration sits on the critical path because of its dependency on the update to the model weights produced at the end of the current iteration. Our study (\S~\ref{sec:insight_1}) allows us to break this dependency: we find that the exploration phase can use the stale model weights from the previous iteration while preserving the relative reward ranking among seeds (Fig.~\ref{fig:heatmap}). 
This allows us to offload seed exploration onto otherwise idle spot GPUs that already have the model weights from the previous iteration, while the reserved cluster runs the training phase. The selected seeds then drive rollout under the updated model in the next iteration. This delivers three benefits: (1) preserving on-policy RL post-training semantics, (2) retaining the faster convergence from seed exploration, and (3) eliminating spot GPU idleness during training.


\PHM{Insight 2. SP reconfiguration can reuse existing state.} The intermittent availability of spot GPUs forces SP groups to reconfigure frequently. The reconfiguration overhead is dominated by two types of delays: CPU scheduler initialization and remote model loading. Both can be mitigated by reusing on-node state: the scheduler persists across reconfigurations, and new workers copy weights locally from a co-located peer with identical weights. This observation reduces SP reconfiguration from minute-scale reboots to sub-second GPU worker launches.


Based on these two insights, we present \SystemName{}, a system that harvests spot GPUs for DiT RL post-training. Given \textbf{Insight 1}, \SystemName{} offloads seed exploration onto the idle spot pool during training. To fit the exploration phase within the training window and maximize exploration quality, \SystemName{} formulates exploration planning as a multi-armed bandit problem and dynamically selects the configuration maximizing reward variance within the time budget given the available spot GPUs.

Given \textbf{Insight 2}, \emph{elastic sequence parallelism} reconfigures SP groups on the fly as spot GPUs leave or rejoin the pool. \SystemName{} decouples the CPU scheduler from GPU workers, keeping it resident on each node across SP changes so its initialization cost is paid only once. When a new GPU worker launches, it copies model weights from a co-located peer rank over NVLink instead of pulling from a remote node, reducing SP reconfiguration to the cost of launching a few GPU worker processes. 

Besides these two components, \SystemName{} also adapts a \emph{preemption-aware pull-based request scheduler}, which lets rollout workers pull requests from a centralized queue and commits the ongoing request state to the reserved cluster for later recovery upon preemption. This design naturally balances load across heterogeneous workers and minimizes progress lost induced by spot GPU preemption.

We implement \SystemName{} on top of the open-source RL platform ROLL~\cite{roll} and evaluate it on Qwen-Image post-training with multiple datasets and image resolutions. Our evaluation shows that \SystemName{} reaches the same target validation score by up to $4\times$ faster, decreasing the total cost by $1.4$-$3.6\times$ compared to state-of-the-art systems~\cite{rlboost} with spot GPUs and $1.9$-$6.4\times$ compared to systems without spot GPUs. This is achieved through both faster training convergence and better spot GPU utilization. This cost reduction does not sacrifice image quality: \SystemName{} achieves superior validation scores on all datasets because dynamic seed exploration effectively selects random seeds that produces more valuable training samples. 

In summary, we make the following contributions:
\begin{itemize}[leftmargin=0cm,itemindent=.4cm,labelwidth=\itemindent,labelsep=0cm,align=left, partopsep=0pt,itemsep=2pt,parsep=0pt]
    \item We identify two insights that enable efficient spot utilization for DiT RL: exploration tolerates stale model weights, and SP reconfiguration can reuse existing states.
    \item We design \SystemName{}, the first system that harvests spot GPUs for DiT RL post-training, around three techniques: spot-side seed exploration, elastic sequence parallelism, and a preemption-aware request scheduler.
    \item We implement \SystemName{} atop ROLL~\cite{roll}. \SystemName{} reaches the same target validation score up to $4\times$ faster than the state-of-the-art baselines, decreasing the total cost by $1.4$-$6.4\times$ with higher validation scores.
\end{itemize}

\section{Background}

\subsection{Diffusion Models and DiT RL Post-Training}
\begin{figure}
    \centering
    \includegraphics[width=\linewidth]{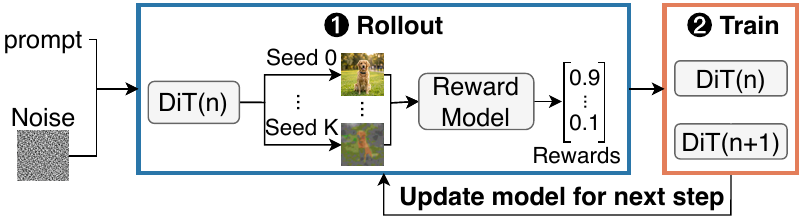}
    \caption{One training step of GRPO-style DiT \posttrain. Each prompt is
repeated $K$ times with different random seeds, and the generated samples are scored by a reward model and used to compute a GRPO gradient update. The next iteration starts after the previous iteration updates the model weights.} 
    \label{fig:grpo}
    \vspace{-10pt}
\end{figure}

Diffusion models generate high-quality images through iterative denoising. A \emph{Diffusion Transformer} (DiT), such as Qwen-Image~\cite{qwen_image}, starts from random Gaussian noise and refines a latent representation over a fixed number of denoising steps (typically 10--50), each requiring a full forward pass through the DiT backbone. The computational cost is thus dominated by these iterative forward passes, and for high-resolution or video generation, sequence-parallel inference across multiple GPUs becomes necessary.

Reinforcement learning (RL) is an effective way to align diffusion models with human preferences and specific generation goals. GRPO-style on-policy DiT RL training uses a three-phase iterative process, shown in Fig.~\ref{fig:grpo}.
\emph{(1) Rollout.} At step $n$, the current model DiT$(n)$ generates $K$
samples from each of $P$ prompts by running denoising with $K$ distinct noise
seeds, producing a total of $P \times K$ rollouts.
\emph{(2) Reward.} Each sample is scored by a reward model, yielding a
per-sample scalar reward $r_k$.
\emph{(3) Training.} The rewards are aggregated into a GRPO policy-gradient objective; a forward and backward pass is performed on DiT$(n)$ to compute the parameter update, and the weights are then broadcast to all rollout workers for the next iteration.
Key parameters of GRPO-style DiT \posttrain{} include image resolution, number of denoising steps, \emph{samples per prompt} $K$, and \emph{number of prompts} $P$, all of which directly impact computational cost across both rollout and training phases.

Due to its compute-intensive nature, GRPO-style DiT \posttrain{} requires thousands of high-end GPUs, which is expensive. For instance, training a commercial-level video DiT requires 4{,}160 GPU days on H200 GPUs at a cost of ~\$200k~\cite{opensora}, and frontier labs are reported to dedicate thousands of H800 GPUs to post-training their video models~\cite{stepvideo}. Research efforts have focused on improving training convergence and GPU utilization to mitigate this high cost.

\subsection{Seed Exploration on DiT RL Training}
Recent algorithmic works~\cite{dancegrpo,treegrpo,branchgrpo,sol-rl} propose \emph{seed exploration} to improve the training convergence. Specifically, they identified that generating samples with high reward contrast, a mix of high-reward and low-reward outputs, provides more informative training signals.
DanceGRPO~\cite{dancegrpo} exploits this by generating $N \gg K$ samples per prompt and selecting the top and bottom $K/2$ by reward to form the training batch, at the cost of proportionally higher rollout latency. TreeGRPO and BranchGRPO~\cite{treegrpo,branchgrpo} use a similar approach, but try to identify high-contrast seeds at the early denoising steps using a tree-based or branch-based search strategy, respectively.
Sol-RL~\cite{sol-rl} proposes first running a cheap exploratory rollout \emph{e.g.}, with reduced denoising steps or a quantized model, to identify high-contrast seeds, then performing full-quality rollout only on the selected seeds.
All approaches can be understood as prepending an \emph{exploration phase} to the standard rollout, which improves convergence by using only high-contrast seeds for training, but at the expense of substantially higher per-iteration latency since exploration adds compute to the critical path.

\subsection{RL Training on Spot GPUs}
\label{sec:background_spot}

The substantial cost of RL post-training has motivated extensive research on
leveraging spot GPUs to reduce resource
expenses. Systems like RLBoost~\cite{rlboost} demonstrate that offloading rollout
generation to preemptible GPUs can significantly reduce LLM-based RL training
costs. Spot GPUs expose spare datacenter
capacity at 71.5\% lower cost than reserved GPUs~\cite{aws_price_list_api,aws_spot_pricing,gcp_spot_pricing,azure_retail_prices_api}, yielding 3.5$\times$ cost savings.
These resources are provisioned to users as cloud VMs~\cite{aws_spot,gcp_spot} or allocated at per-GPU granularity by a centralized cluster scheduler~\cite{gfs,pollux}. The number of GPUs assigned to a job  can be reclaimed
unpredictably with only brief grace periods (30--120 seconds depending on the
provider~\cite{aws_grace,gcp_grace}), requiring workloads to checkpoint or accept progress loss. Additionally, spot capacity is fragmented across nodes~\cite{gpufragmentation,weng2023beware}: the training phase requires high-speed interconnects such as NVLink, which are often limited to intra-node communication in cloud deployments, but spot GPUs are scattered across many nodes with only slower inter-node links, making them well-suited for embarrassingly parallel rollout but not for tightly-coupled training.
This leads to a severe problem: allocated spot GPUs sit idle during the training phase with no useful work to do, wasting a large portion of preemptible resources. Existing works targeting LLM-based post-training mitigate such idleness by pipelining rollout with training, which relies on the long-tail distribution of LLM rollout latencies~\cite{rlboost}.
However, such characteristics do not hold for DiT RL, where rollouts complete in near-uniform time, eliminating overlap opportunities and leaving spot GPUs idle during training, resulting in significant resource waste and higher cost.

\section{Motivation}
\label{sec:motivation}

In this section, we identify the unique challenges through charactering the workload of DiT RL post-training and propose two insights
in response to these challenges.

\subsection{Challenges}

To understand how spot GPUs affects different phases of post-training pipeline, we characterize the workload using Qwen-Image (20B parameters) on 4 reserved GPUs for training and rollouts and up to 12 H100 spot GPUs dedicated to rollout. We configure the training job with image size $512 \times 512$, $N=32$ prompts, $K=16$ samples per prompt, 20 denoising steps for rollout inference. We use vLLM-omni~\cite{vllm_omni} for rollouts and FSDP~\cite{fsdp} for training, measuring performance across five consecutive training iterations. Following mainstream practice~\cite{rollart}, we deploy the reward model as a service that scores samples asynchronously during rollout with negligible overhead; we therefore omit reward time from the breakdown. We measure the phase-wise time breakdown across each training iteration, tracking rollout and training phases. 

\begin{figure}
    \centering
    \includegraphics[width=\linewidth]{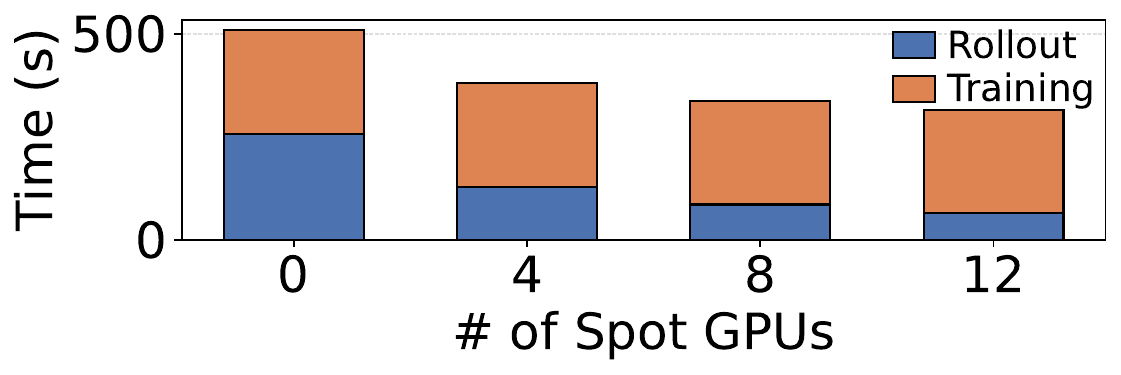}
    \vspace{-20pt}
    \caption{Per-step time breakdown vs.\ number of spot GPUs. Adding spot GPUs significantly reduces rollout latency (up to $4\times$), while training time remains unchanged since it requires stable reserved GPUs. Reward scoring is deployed as an asynchronous service and omitted from the breakdown due to negligible overhead.
    }
    \vspace{-10pt}
    \label{fig:breakdown}
\end{figure}

We start by characterizing the major bottleneck of DiT RL \posttrain{} under different numbers of spot GPUs. We compare the per-iteration time of \posttrain{} with and without spot GPUs. On 4 reserved GPUs alone, rollout and training each take around half of the iteration time. When offloading rollout to spot GPUs, rollout latency scales nearly linearly: adding 4, 8, and 12 spot GPUs reduces it by $2\times$, $3\times$, and $4\times$ respectively (Fig.~\ref{fig:breakdown}), while training time remains constant since it requires stable reserved GPUs. This near-linear scalability makes the rollout phase of DiT RL particularly well-suited for spot GPUs. However, naively applying spot GPUs introduces two unique challenges that existing LLM-based systems are not designed to address.

The first challenge is that spot GPUs remain idle during the training phase. We evaluate the utilization of the spot GPUs in conventional deployment, similar to~\cite{rlboost}. We allocate spot GPUs exclusively for rollout and reserve stable on-demand GPUs for training. In our experiments, spot GPUs are idle for 47\% of iteration time, i.e., during the entire training phase. In contrast to LLM model rollouts~\cite{rlboost} that can take variable time to complete, diffusion rollouts go through a fixed number of denoising steps, leaving no opportunity to overlap them with subsequent training phases (\S\ref{sec:background_spot}).

\begin{figure}
    \centering
    \includegraphics[width=\linewidth]{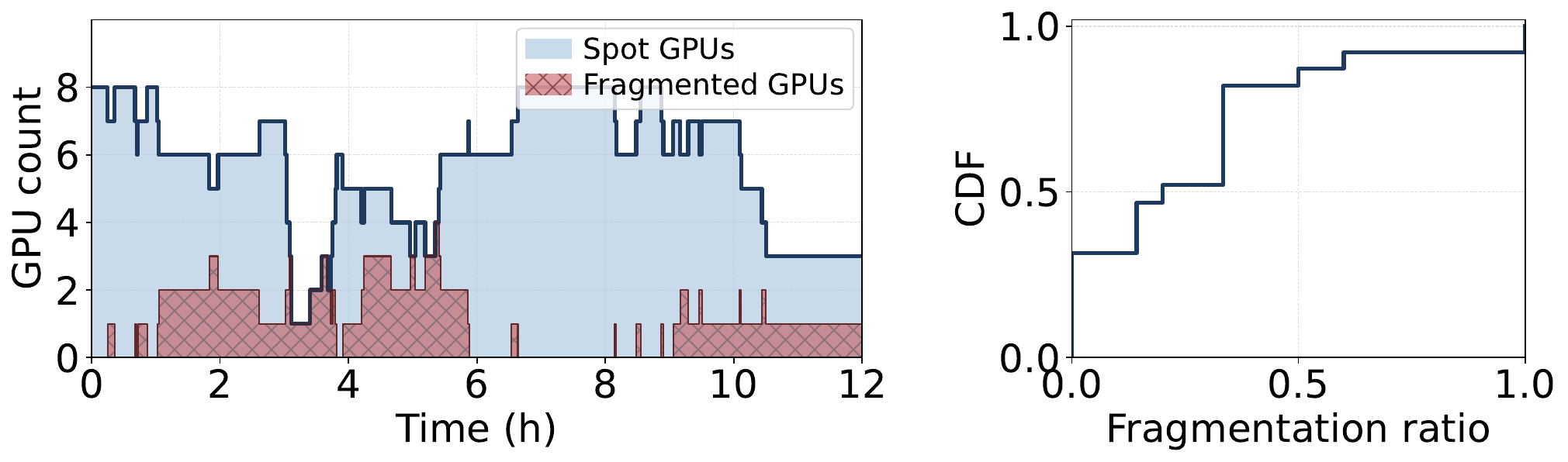}
    \vspace{-20pt}
    \caption{
    Spot GPU fragmentation under spot GPU dynamics (RLBoost trace~\cite{rlboost}, 4 nodes $\times$ 2 H100 GPUs, SP$=2$). 
\emph{Left}: available spot GPUs and fragmented GPUs over time. 
\emph{Right}: time-weighted CDF of the fragmentation ratio 
(fragmented / total available GPUs). 
    }
    \vspace{-20pt}
    \label{fig:spot_fragmentation}
\end{figure}

The second challenge is that spot preemptions fragment GPU topology and break SP groups. Large diffusion models often require multi-GPU sequence parallelism (SP) to fit model parameters within memory constraints. When any GPU in an SP group is preempted, the entire group collapses and in-flight rollouts must be discarded. In this work, we assume spot resources are allocated at per-GPU granularity by a centralized cluster scheduler~\cite{gfs,pollux}, where the number of GPUs assigned to a job may change during execution, and define a GPU as \emph{fragmented} when its node cannot host a complete SP group. For example, in the $\mathrm{SP}{=}2$ setting, GPU fragmentation means that only one GPU remains on a node, hence the desired group cannot be hosted. We conduct a trace-driven study using a production preemption trace~\cite{thorpe2023bamboo} on a cluster with 8 spot GPUs across 4 nodes ($\mathrm{SP}{=}2$). This trace lacks GPU placement information; hence, we assume uniform GPU arrivals and revocations across nodes. As shown in Fig.~\ref{fig:spot_fragmentation}, fragmentation is prevalent: across 70\% of the trace duration, at least one GPU is fragmented, and in over 50\% of the time, more than 20\% of available spot GPUs cannot participate in SP inference due to incomplete node-level groups. Naively reconstructing the SP group with an updated set of GPUs introduces prohibitive overhead: loading weights and reinitializing the inference engine take up to 2 minutes for a 20B model.

\label{sec:motivation-example}

\subsection{Insights}
We analyze the root cause of these challenges and identify two insights that guide our system design.

\subsubsection{Insight 1: Exploration Can Tolerate Stale Model Weights}
\label{sec:insight_1}
The root cause of both challenges above is the serial dependency in the GRPO algorithm: both exploration and rollout must be conducted on the latest model weights, so they cannot proceed until the training phase finishes and produces an update. This dependency leaves spot GPUs idle during training and forces exploration onto the critical path. We observe that these two problems can resolve each other: if exploration can tolerate stale model weights (i.e., weights from the previous iteration), we can offload it to spot GPUs during training, simultaneously harvesting idle resources and moving exploration off the critical path.

\begin{figure}
    \centering
    \begin{subfigure}[b]{0.49\linewidth}
        \includegraphics[width=\linewidth]{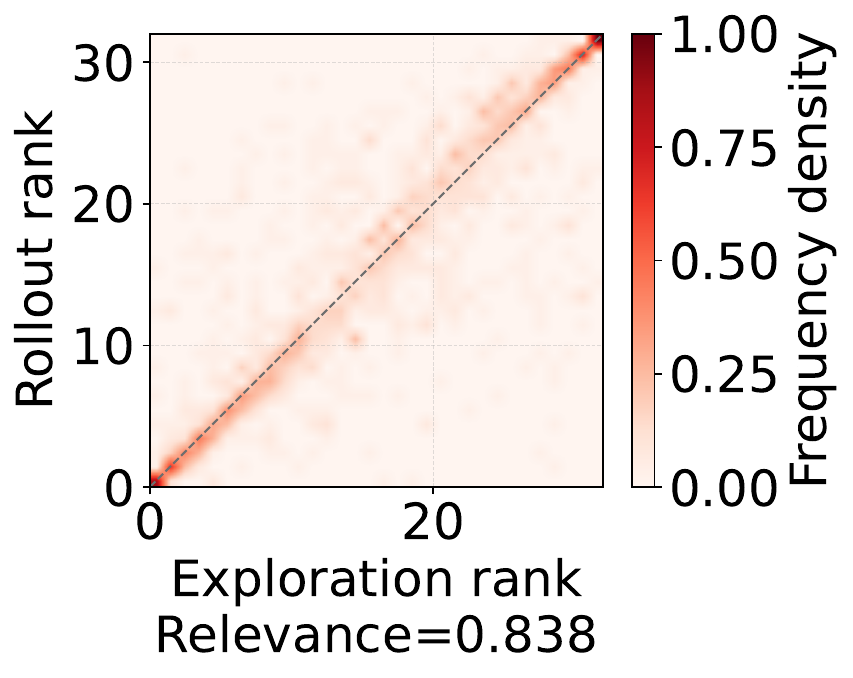}
    \caption{DeepSeek-OCR.}
    \label{fig:heatmap_ocr}
    \end{subfigure}
    \begin{subfigure}[b]{0.49\linewidth}
        \includegraphics[width=\linewidth]{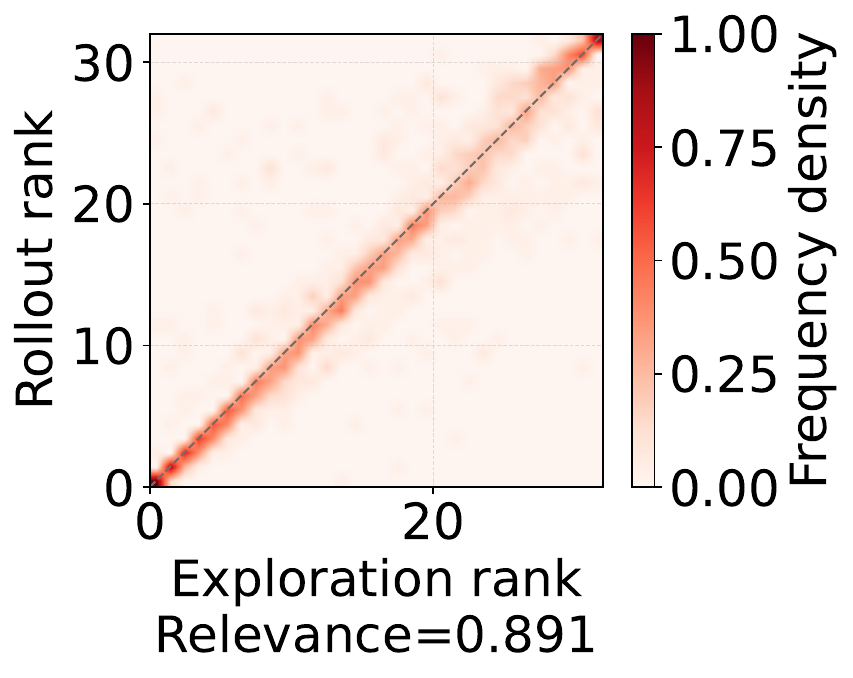}
    \caption{Geneval.}
    \label{fig:heatmap_geneval}
    \end{subfigure}
    \vspace{-10pt}
    \caption{
Heat maps comparing reward ranks from stale-model exploration and updated-model rollout using the same prompts and random seeds. Each cell shows how often a seed ranked $i$ under the previous model is ranked $j$ under the updated model; darker regions indicate higher frequency. The strong diagonal pattern shows that stale-model exploration preserves intra-group reward ranking.
}
\vspace{-10pt}
    \label{fig:heatmap}
\end{figure}

To validate this hypothesis that seed rankings can be reliably predicted with stale weights, we conduct the following study.
We collect five pairs of consecutive checkpoints from a full training run on the entire dataset. Each pair consists of a stale model (before update) and the corresponding updated model (after update). For each prompt, both models generate 32 samples using the same set of random seeds, and the samples are ranked by reward within the 32-seed group. We compare the reward ranking from the stale model (\emph{Exploration rank}) with the ranking from the updated model (\emph{Generation rank}). 
Fig.~\ref{fig:heatmap} plots the distribution of Generation rank versus Exploration rank for two datasets. Each cell represents how often a seed ranked $i$ under the stale model is ranked $j$ under the updated model, with darker cells indicating higher frequency. For both datasets, the distribution is strongly concentrated along the diagonal, confirming that stale-model exploration preserves the intra-group seed ranking by reward.

This finding enables us to offload seed exploration to spot GPUs using stale weights while the reserved cluster performs training. The selected high-contrast seeds are then used for full-quality rollout with the updated model in the following iteration, ensuring that training consumes on-policy samples with larger reward variance while utilizing otherwise idle spot resources.

\subsubsection{Insight 2: SP Reconfiguration Can Reuse Existing State.}
\label{sec:insight_2}
\begin{figure}
    \centering
    \includegraphics[width=\linewidth]{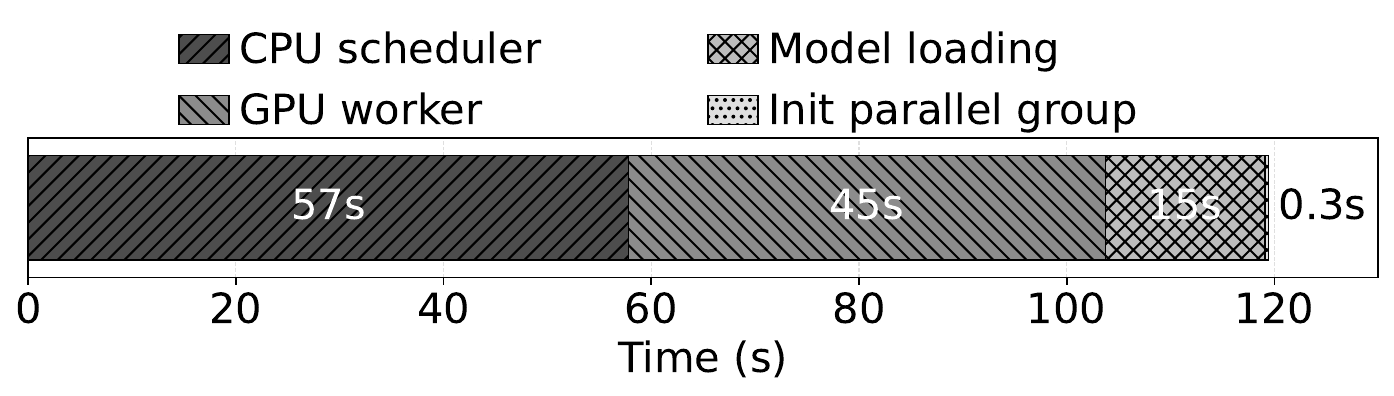}
    \vspace{-20pt}
    \caption{Initialization time breakdown of a typical DiT inference engine. CPU scheduler initialization and remote model weight loading dominate the startup latency, while GPU worker launch and communication group setup take only a small fraction.}
    \label{fig:init-breakdown}
    \vspace{-10pt}
\end{figure}

The root cause of prohibitive SP reconfiguration overhead lies in the assumption that the entire inference engine must be restarted when the SP degree changes, which brings a 2-minute delay as shown in Fig.~\ref{fig:init-breakdown}. 
To pinpoint where the actual cost goes, we profile the initialization time of a typical inference engine (vLLM-Omni~\cite{vllm_omni}) used in rollout phase of DiT \posttrain. We load model weights from a remote node over a 50Gbps Ethernet link
to emulate the weight synchronization 
during the training phase. As shown in Fig.~\ref{fig:init-breakdown}, the startup latency is dominated by two components: CPU scheduler initialization and remote model weight loading, which together account for over 62\% of the total time. In contrast, GPU worker process launch and communication parallel group setup take only 38\% of the total time.

Neither of these dominant costs is inherent to changing SP degree. The CPU scheduler’s request-level state (queues, metadata, denoising step counters) is independent of the number of GPU workers. Within an SP group, all GPU workers share identical model weights, so a new worker can copy weights from a co-located peer over NVLink instead of fetching them remotely. As a result, SP reconfiguration shrinks from minute-scale engine restarts to simply launching a few GPU workers and rebuilding the communication group, allowing the system to track spot availability with minimal disruption.

\section{System Design}

\label{sec:system-approach}


To systematically mitigate the performance bottlenecks identified in \S\ref{sec:motivation}, we structure \SystemName{} around three core design principles. First, intra-group reward rankings remain highly robust under weight updates (\textbf{Insight~1}), thus we employ \emph{spot-side seed exploration} to offload candidate screening to volatile spot GPUs during the training phase. To overlap the exploration latency with the training phase, \SystemName{} dynamically optimizes hyper-parameters via an online bandit framework. Second, sequence-parallel reconfiguration overhead stems from redundant initialization rather than physical layout changes (\textbf{Insight~2}). To eliminate this overhead, we introduce \emph{elastic sequence parallelism} that preserves the initialized model state and reshapes only the active worker group via intra-node weight broadcasts, rather than costly engine restarts. Third, spot capacity is inherently transient; thus, we use \emph{preemption-aware persistence} to commit intermediate rollout state to stable, reserved infrastructure, ensuring the system tolerates sudden reclamation without discarding completed work.

\subsection{\SystemName{} Architecture}
\label{sec:architecture}

\begin{figure}
    \centering
    \includegraphics[width=\linewidth]{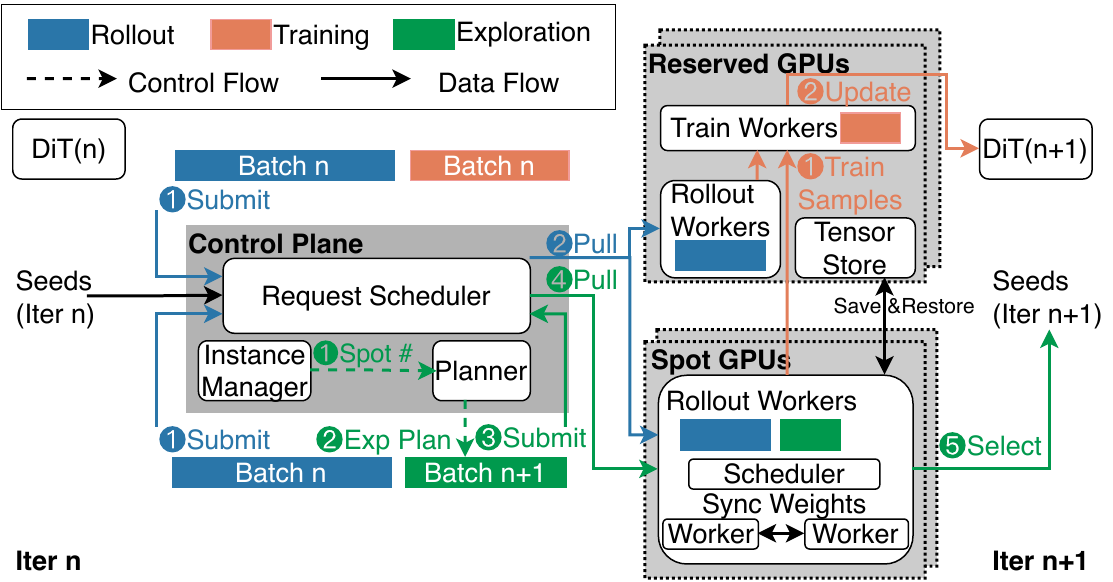}
    \vspace{-10pt}
    \caption{System architecture of \SystemName{}. Colors distinguish the phases of DiT \posttrain. Dashed arrows represent control flow, solid arrows represent data flow. Circled numbers indicate the execution order within each phase.
    }
    \vspace{-20pt}
    \label{fig:design}
\end{figure}

Fig.~\ref{fig:design} shows the architecture of \SystemName{}. The system manages two pools of GPUs distributed across nodes: a stable set of \emph{Reserved GPUs} that host both training and part of the rollout, and a volatile set of \emph{Spot GPUs} that elastically scale rollout and exploration capacity. While intra-node GPUs are tightly coupled via high-bandwidth NVLink interconnects, cross-node communication between the reserved and spot pools relies on slower commodity Ethernet, thus introducing a severe bandwidth asymmetry that directly shapes our data plane design. The control plane, hosted on the reserved side, contains a \emph{Request Scheduler} that maintains a centralized request queue for both on-policy rollout and exploration requests, a \emph{Planner} that determines the exploration configuration at each iteration boundary, and an \emph{Instance Manager} that tracks the current number and lifecycle of spot GPUs. The data plane spans both reserved and spot GPUs: \emph{Train Workers} on reserved GPUs perform gradient updates, \emph{Rollout Workers} on both reserved and spot GPUs execute inference requests, and a \emph{Tensor Store} on the reserved node durably holds intermediate rollout state including active denoising latents and request metadata, hence providing a resilient fallback to tolerate sudden resource reclamation. Each spot node runs a local \emph{Scheduler} that coordinates its \emph{Workers} into SP groups and handles weight synchronization upon elastic reconfiguration. Following~\cite{rollart}, we deploy the reward model as an external microservice outside the GPU cluster. Reward scoring runs asynchronously with rollout and exploration, thus entirely removing validation scoring from the critical path and preventing GPU memory or compute contention on the active training GPUs.

\subsection{Asynchronous Iteration Workflow}

We now describe the workflow of a single training iteration (Iter $n$), which consists of three phases.

\begin{enumerate}[leftmargin=0cm,itemindent=.2cm,labelwidth=\itemindent,labelsep=0cm,align=left,topsep=0pt,partopsep=0pt]
\setlength\itemsep{0pt}

\item \textbf{Rollout and Reward Phase}: The current prompt batch (Batch $n$) is submitted to the Request Scheduler (\ding{182}). Both spot and reserved Rollout Workers pull tasks from the centralized queue to execute inference in parallel (\ding{183}). Once all rollout sequences are generated, \SystemName{} asynchronously evaluates the samples, thus completing the phase and splitting the remaining iteration compute into two concurrent paths.


\item \textbf{Training Phase}: All Rollout Workers stream the generated training samples to the Train Workers (\ding{182}). The Train Workers perform the gradient update to produce the next-generation model weights, denoted as DiT($n+1$) (\ding{183}). Cross-cluster communication runs over a 50\,Gbps Ethernet network, thus broadcasting these updated parameters to the volatile Spot GPUs requires approximately 15\,s (Fig.~\ref{fig:elastic_sp}); 

\item  \textbf{Exploration phase.} This phase runs concurrently with training, leveraging \textbf{Insight~1} that exploration tolerates stale weights (\S\ref{sec:insight_1}). \ding{182} The Instance Manager reports the current spot GPU count to the Planner. \ding{183} The Planner uses this count together with offline-profiled per-request latencies to determine an exploration plan that fits within the training window (\S\ref{sec:explore_on_spot}). \ding{184} The Planner submits the next iteration's prompts (Batch $n$+1) to the Request Scheduler as exploration requests. \ding{185} Only spot Rollout Workers pull these exploration requests and execute them. \ding{186} After execution, a seed selection step picks the best seeds based on reward scores for the next iteration. With both the updated DiT($n$+1) and the selected seeds ready, the system proceeds to Iter $n$+1.

\end{enumerate}

We describe how \SystemName{} handles spot instance preemption (\S\ref{sec:step-wise-scheduler}) and elastic reconfiguration of SP groups (\S\ref{sec:elastic_sp}) in the following subsections.

\subsection{Dynamic Exploration via Bandit-Based Hyperparameter Adaptation}
\label{sec:explore_on_spot}

As described in \S\ref{sec:architecture}, the Exploration Phase runs concurrently with training on spot GPUs. However, the available spot capacity fluctuates across iterations; if exploration exceeds this budget, the unfinished portion falls onto the critical path of the subsequent iteration, thereby directly increasing per-step latency. The Planner must therefore determine, at the beginning of each exploration phase, an appropriate \emph{exploration configuration} that maximizes reward variance within this variable budget. We apply an online bandit-based adaptation mechanism that dynamically adjusts these parameters throughout training.

\subsubsection{Action Space.}
We formulate the per-iteration exploration configuration as a multi-armed bandit problem over a discrete action space. Each action is parameterized by two axes that most directly affect exploration cost and quality: the number of sequences $d$ (rollout sequences generated per prompt) and the effective denoising steps $s$ per sequence. However, directly reducing the number of denoising steps degrades generation quality and distorts reward signals, undermining exploration accuracy. We therefore control effective denoising steps via a caching strategy rather than truncation.

Specifically, we use \tcache~\cite{teacache}, a caching technique for diffusion models that caches and reuses intermediate transformer outputs across adjacent denoising steps when their residual difference falls below a threshold. Each \tcache{} threshold corresponds to an average number of effective denoising steps $s$, which we obtain via offline profiling for each dataset. Let $S$ denote the total scheduled denoising steps under full computation. We define the action space over the two axes as $a = (d, s)$, where $s \in \{s_1, s_2, \ldots, s_K\}$ is drawn from the set of profiled effective step counts. This formulation naturally extends to richer configuration spaces by adding additional axes to the action tuple. 
We verify in \S\ref{sec:sensitivity} that reduced denoising steps via caching preserve exploration accuracy; crucially, this adaptation is applied strictly during seed screening, whereas on-policy training updates consistently utilize full-fidelity sequences to maintain gradient integrity.

Given these two axes, we construct the action space by retaining only the configurations that fit within the exploration time budget. Let $T_{\text{train}}$ denote the measured model update duration per iteration, $N_{\text{spot}}$ the number of spot GPUs reported by the Instance Manager, and $t_{\text{step}}$ the profiled per-step latency. The aggregate exploration time window is $W = T_{\text{train}} \cdot N_{\text{spot}}$. For a fixed number of prompts $C$, the planned time of an action $a = (d, s)$ is
\[
T_{\text{plan}}(a) = d \cdot C \cdot s \cdot t_{\text{step}}.
\]
The eligible action space is then defined as
\[
\mathcal{A} = \bigl\{\, (d, s) \;\bigm|\; T_{\text{plan}}(d, s) \leq W \,\bigr\},
\]
ensuring that all retained configurations can complete exploration within the time window afforded by model training on the Reserved GPUs. The maximum number of sequences and the minimum number of denoising steps
jointly determine the size of the action space, and we select these values
through a sensitivity study in \S~\ref{sec:sensitivity}.

\subsubsection{Bandit Feedback.}
To evaluate the effectiveness of an exploration configuration, we need a feedback signal that measures how much a given action improves reward variance compared to the natural variance of the current policy. The target policy transitions continuously across successive training iterations; thus, the baseline variance shifts over time, rendering absolute reward variance an unreliable indicator of exploration quality. We therefore reserve a small subset of prompts (4 per iteration) as an unexplored group that use the default configuration, providing a per-iteration reference for the current policy's intrinsic variance.

Concretely, after each iteration we compute the per-prompt reward standard deviation $\sigma_g$ for each prompt group $g$ in the training batch. Let $\mathcal{G}_{\text{cov}}$ and $\mathcal{G}_{\text{unc}}$ denote the sets of explored and unexplored prompt groups, respectively. We compute:
\[
\bar{\sigma}_{\text{all}} = \frac{1}{|\mathcal{G}_{\text{cov}}| + |\mathcal{G}_{\text{unc}}|} \sum_{g \in \mathcal{G}_{\text{cov}} \cup \mathcal{G}_{\text{unc}}} \sigma_g, \qquad
\bar{\sigma}_{\text{unc}} = \frac{1}{|\mathcal{G}_{\text{unc}}|} \sum_{g \in \mathcal{G}_{\text{unc}}} \sigma_g.
\]
The bandit feedback is then defined as the ratio
\[
r = \frac{\bar{\sigma}_{\text{all}}}{\bar{\sigma}_{\text{unc}}},
\]
which captures the degree to which exploration amplifies reward variance relative to the baseline variance of the current policy. A higher ratio indicates that the chosen configuration produces rollouts with greater reward contrast, providing stronger learning signals for policy optimization.

\subsubsection{UCB-based Action Selection.}
We adopt the Upper Confidence Bound (UCB) strategy to balance exploitation of well-performing configurations with exploration of under-sampled ones. For each action $a_i \in \mathcal{A}_{\text{elig}}$, we maintain a sliding window of the $W_b$ most recent bandit feedback observations $\{r^{(1)}_i, \ldots, r^{(n_i)}_i\}$, where $n_i = \min(N_i, W_b)$ and $N_i$ is the total number of times action $a_i$ has been selected. Let $N = \sum_{i} N_i$ denote the total number of bandit updates. The empirical mean feedback is 
\[
\hat{\mu}_i = \frac{1}{n_i} \sum_{j=1}^{n_i} r^{(j)}_i.
\]
The UCB score for action $a_i$ is then
\[
\text{UCB}(a_i) = \hat{\mu}_i + \beta \sqrt{\frac{\ln(N+1)}{n_i}},
\]
where $\beta > 0$ is the exploration coefficient that controls the trade-off between exploitation and exploration: a larger $\beta$ encourages the Planner to try under-sampled configurations more aggressively, while a smaller $\beta$ favors configurations with proven high reward. We choose $\beta$ that can effectively search the action space while converging to a stabilized set of hyper-parameters in early stage of \posttrain{} through a sweep study in \S\ref{sec:beta}.   For actions that have not yet been observed ($n_i = 0$), we set $\text{UCB}(a_i) = +\infty$ to ensure they are prioritized. At each iteration, the Planner selects
\[
a^* = \arg\max_{a_i \in \mathcal{A}_{\text{elig}}} \text{UCB}(a_i),
\]
with ties broken in favor of lower planned cost, fewer effective denoising steps, and fewer sequences, respectively.

\subsubsection{Handling Incomplete Exploration.}
Despite the dynamic adaptation of exploration parameters, the inherent unpredictability of spot GPU availability may still lead to mid-iteration resource changes, causing exploration to fall short of completion before the training phase ends. In such cases, we pause exploration and proceed to complete the model update using all available GPUs. After the update, the remaining exploration workload is resumed and executed using all Rollout Workers on both Reserved GPUs and Spot GPUs. This design ensures stable exploration coverage across iterations, although at the cost of potential additional overhead, which we analyze in the evaluation section.

\subsection{Elastic Sequence Parallelism}
\label{sec:elastic_sp}

Spot fragmentation forces \SystemName{} to change Rollout Workers' SP degree as GPUs leave and rejoin. As shown in \S\ref{sec:insight_2}, the dominant SP reconfiguration costs—CPU scheduler initialization and remote weight loading—can be removed by reusing on-node state. \SystemName{} leverages this with two mechanisms.

\subsubsection{Decoupled Persistent Scheduler.}
Guided by this breakdown, \SystemName{} decouples the local Scheduler from the GPU Workers and keeps the Scheduler resident on the spot node across SP changes. The persistent Scheduler retains all request-level state, including queued prompts, in-flight requests, denoising step counters, and request metadata, so that its initialization cost is paid only once per node lifetime. When spot availability changes, \SystemName{} kills or launches only the affected Workers and rebuilds the communication group around the new rank set. The Scheduler then resumes dispatching requests to the reshaped worker pool without losing in-flight rollouts, which removes the dominant CPU scheduler initialization cost from the SP reconfiguration path.

\subsubsection{Intra-Node Weight Loading.}
Launching a fresh Worker still requires populating its weights, and loading from a remote node would reintroduce the inter-node transfer shown in Fig.~\ref{fig:init-breakdown}. \SystemName{} avoids this path whenever a peer rank from the same SP group is already running on the same physical node: the new Worker copies model weights from that peer over NVLink instead of pulling them across the network. Because ranks in an SP group share identical weights, the intra-node copy is equivalent to a remote load and needs no additional synchronization with the training cluster. \SystemName{} falls back to remote loading only when no co-located peer is available. Together with the persistent Scheduler, intra-node weight loading reduces SP reconfiguration to the cost of launching a small number of Workers. The rollout pipeline still pauses briefly during this transition, but the stall is bounded by the residual worker-side work rather than by full engine re-initialization, allowing \SystemName{} to track spot availability with limited disruption.

\subsection{Preemption-aware Request Scheduler}
\label{sec:step-wise-scheduler}

\SystemName{} adopts a preemption-aware pull-based Request Scheduler to handle both heterogeneous worker speeds and unpredictable spot revocations. Unlike traditional DiT post-training, \SystemName{} introduces an additional exploration phase and executes rollouts on volatile spot resources. Statically assigning requests to Rollout Workers at the beginning of each iteration easily leads to imbalance: workers with different SP degrees finish at different speeds, and spot workers may be reclaimed before completing their assigned requests. The Request Scheduler therefore maintains a centralized request queue and lets each Rollout Worker pull new requests when it becomes available. This pull-based design adapts to heterogeneous throughput and avoids overloading slow or unstable workers.

To tolerate preemption, \SystemName{} maintains the Tensor Store on reserved nodes that holds intermediate rollout state, including latent tensors and scheduler metadata. When a spot Worker is about to be reclaimed, the Instance Manager detects the upcoming termination and instructs the Worker to stop accepting new requests and commit its in-flight rollout state to the Tensor Store. Once the commit completes, the unfinished request is returned to the Request Scheduler and resumed by another available Rollout Worker, preserving partial progress and reducing redundant computation. If a spot GPU is hard-killed before the commit finishes, the Request Scheduler detects the unfinished request through lifetime monitoring and re-enqueues it for full re-execution by another Worker.

\section{Implementation}
\label{sec:implementation}
We implement \SystemName{} on top of ROLL~\cite{roll} in approximately 6k lines of Python. The system uses vLLM-omni~\cite{vllm_omni} as the DiT rollout inference engine and FSDP2~\cite{fsdp} for training. The algorithm backbone is FlowGRPO~\cite{flowgrpo}. 

\PHM{Dynamic Exploration.} 
For seed selection in the bandit-based exploration planner~(\S\ref{sec:explore_on_spot}), we adopt a top-$k$/bottom-$k$ strategy over per-prompt exploration rewards, retaining the highest- and lowest-scoring seeds to construct a seed bank that maximizes intra-group reward diversity for subsequent iterations.

\PHM{Elastic Sequence Parallelism.}
The CPU scheduler and GPU workers communicate through ZeroMQ~\cite{zmq} shared-memory message queues, making the scheduler agnostic to the live worker count. 
When spot capacity changes, the scheduler triggers SP group reinitialization as described in \S\ref{sec:elastic_sp}. For intra-node weight transfer, parameter metadata is announced over the existing broadcast queue, and tensors are transferred via \texttt{torch.distributed.\_broadcast\_coalesced} rooted at the source rank over the rebuilt NCCL group.

\PHM{Preemption-aware Request Scheduler.}
We implement the preemption-aware Request Scheduler described in \S\ref{sec:step-wise-scheduler}. The Tensor Store is built on Mooncake Store~\cite{qin2024mooncake}. Each request follows a state machine (Pending $\rightarrow$ In-flight $\rightarrow$ Done/Recompute/Aborted) that tracks its lifecycle through dispatch, execution, and potential preemption recovery.

\section{Evaluation}
\label{sec:evaluation}

\subsection{Methodology}

\PHM{Hardware Setup.} We use 4 H100 GPUs on a reserved node and 8 H100 GPUs on 4 spot nodes, with 2 GPUs per spot node. GPUs within the same node are connected by NVLink with 900GB/s bandwidth. Each spot node has a 50Gbps Ethernet vNIC, and the reserved node has a 200Gbps Ethernet NIC for communicating with spot nodes. 

We colocate training workers and rollout workers on the reserved node and deploy only rollout workers on the spot nodes. Similar to prior work~\cite{rollart}, we deploy the reward model as a service outside this cluster. Rollout and exploration workers interact with it through the network to send samples and receive scores asynchronously. In the following study, we ignore reward latency, as it's off the critical path.

\PHM{Spot Trace.} We use the 12-hour spot availability production trace from Bamboo~\cite{thorpe2023bamboo}, which records the availability of $2\times$ H100 spot GPUs over time. While RLBoost~\cite{rlboost} extracts three 2-hour segments from this trace for evaluation, we use the complete 12-hour trace without segmentation to capture the full temporal dynamics of spot GPU availability. Since the original trace does not record per-node GPU placement, we randomly assign each arrival and revocation event to one of the four spot nodes.

\PHM{Dataset.} We evaluate \SystemName{} on two text-to-image benchmarks: \ocr{} and \geneval. \ocr{} measures text-rendering quality, while \geneval{} evaluates object-level and compositional alignment. For reward scoring, we use Qwen3-VL-8B~\cite{qwen3_vl} on the \ocr{} dataset and a compositional rule-based reward built on Mask2Former~\cite{mask2trans} and OpenCLIP ViT-L/14~\cite{openclip} for the \geneval{} dataset.

\PHM{Model Setup.} We use the Qwen-Image 20B text-to-image model as the base model for all experiments. During rollout, we use 20 denoising steps for the \ocr{} dataset and 10 denoising steps for the \geneval{} dataset, and generate a group of 16 samples for each prompt, same as ~\cite{sol-rl,flowgrpo,dancegrpo}.
We evaluate both datasets at two resolutions: 512$\times$512 (low resolution) and 1280$\times$1280 (high resolution). 
At low resolution, we set the initial SP level of all rollout workers to 1, and at high resolution, we set the initial SP level to 2. To ensure reproducibility, we assign each sample a deterministic random seed, making the candidate seed set for exploration fully reproducible across runs. Additional hyperparameters for convergence evaluation are provided in Appendix~\ref{app:convergence_hyperparams}.

\PHM{Baselines.} We use the following systems as our baselines.
\begin{itemize}[leftmargin=0cm,itemindent=.4cm,labelwidth=\itemindent,labelsep=0cm,align=left, partopsep=0pt,itemsep=2pt,parsep=0pt]
    \item \Bs{} is the state-of-the-art spot-based baseline for DiT \posttrain, implemented on top of ROLL~\cite{roll}. It uses both reserved and spot GPUs during rollout, and only reserved GPUs during training.
    \item \Bes{} is a training framework for multi-modal \posttrain{} including DiT. We enhance it with an exploration phase before rollout that generates $N \gg K$ samples and selects the top and bottom $K/2$ by reward score to form the training batch. We run it with the same worker configuration as \Bs, with spot GPUs enabled.
    \item \B{} is the reserved-only baseline for evaluating cost efficiency without spot GPUs. We provision $3\times$ the reserved GPUs of \Bs(12 reserved GPUs total), matching the maximum available GPUs under a  setup with spot GPUs. It runs both rollout and training on all 12 reserved GPUs.
    \item \Be{} is the reserved-only baseline for \Bes. It prepend an exploration phase in rollout and uses the same hardware configurations as \B.
\end{itemize}

We first present end-to-end cost comparison showing that \SystemName{} reduces total training cost (\S\ref{sec:e2e}). We then evaluate each component individually: dynamic exploration and its effect on convergence speed and overhead (\S\ref{sec:dynamic_explore}), elastic sequence parallelism and its ability to track spot capacity changes in real time (\S\ref{sec:elastic_sp}), and the preemption-aware request scheduler under varying preemption frequencies (\S\ref{sec:preemption}). Finally, we isolate each component's contribution via ablation (\S\ref{sec:ablation}), verify near-linear scaling with spot GPU count (\S\ref{sec:scalability}), and study sensitivity to dynamic exploration's hyperparameters (\S\ref{sec:sensitivity}).

\subsection{End-to-End Experiments \& Cost Reduction}
\label{sec:e2e}

\begin{figure}
    \centering
    \includegraphics[width=\linewidth]{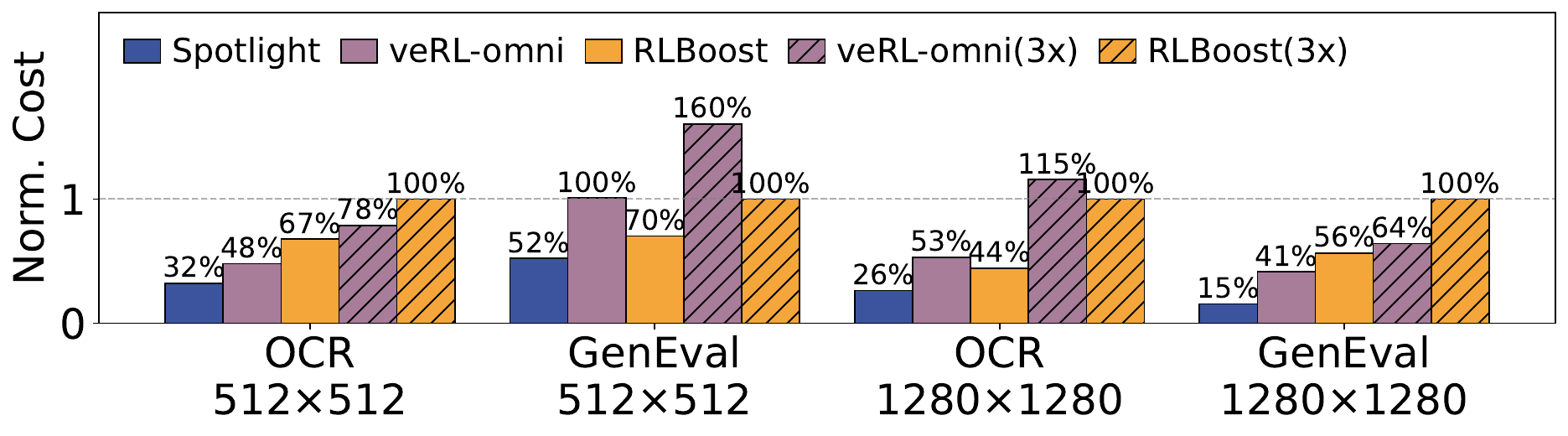}
    \caption{\textbf{[E2E cost]:} Total cost normalized to \B{} across five different setups in end-to-end experiments across two image resolutions and datasets. Hatched bars indicate systems without spot GPUs.
    }
    \label{fig:e2e_cost}
\end{figure}
We run each configuration until its validation score saturates and compare the total monetary cost at the point where each run first reaches the target score. We set the target score to the lowest saturated validation score among all configurations: $0.7$ for \ocr{} $512\times512$, $0.75$ for \geneval{} $512\times512$, $0.6$ for \ocr{} $1280\times1280$, and $0.5$ for \geneval{} $1280\times1280$. We adopt the pricing methodology of~\cite{rlboost} with publicly listed rates from ~\cite{aws_price_list_api,aws_spot_pricing,gcp_spot_pricing,azure_retail_prices_api}: \$2.87/hour for a spot GPU and \$10.08/hour for a reserved GPU. Because the number of active spot GPUs varies over time, we compute spot cost by integrating the instantaneous spot GPU count over time at the per-GPU hourly rate.

Figure~\ref{fig:e2e_cost} compares the normalized monetary cost across all datasets and resolutions, with each group's cost normalized to \B{}. \SystemName{} consistently achieves the lowest cost across all configurations, reducing total cost by $1.35$--$6.39\times$ overall. Compared with the reserved-only baselines \B{} and \Be, \SystemName{} reduces cost by $1.92$--$6.39\times$ because spot GPUs cost only 28\% of reserved GPUs, significantly reducing the total cost. Compared with \Bs, \SystemName{} achieves $1.35$--$2.22\times$ lower cost because dynamic exploration significantly improves training convergence, reducing the number of iterations to reach the target score by up to $4\times$ (\S\ref{sec:dynamic_explore}). Compared with \Bes, \SystemName{} achieves $1.71$--$3.62\times$ lower cost because \SystemName{} overlaps seed exploration with training by utilizing otherwise-idle spot GPU time, while \Bes{} places exploration on the critical path, resulting in $1.32$--$1.54\times$ higher per-iteration latency.

\subsection{Dynamic Exploration: Accuracy \& Convergence}
\label{sec:dynamic_explore}
\begin{figure}
    \centering
    \begin{subfigure}[b]{0.49\linewidth}\includegraphics[width=\linewidth]{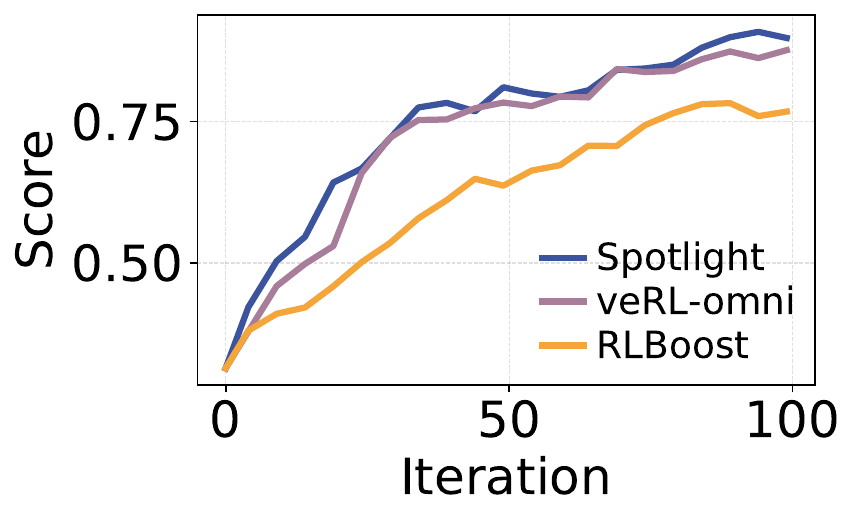}
        \caption{DeepSeek-OCR, $512\times512$.}
        \label{fig:exp_ocr_512}
    \end{subfigure}
    \begin{subfigure}[b]{0.49\linewidth}
        \includegraphics[width=\linewidth]{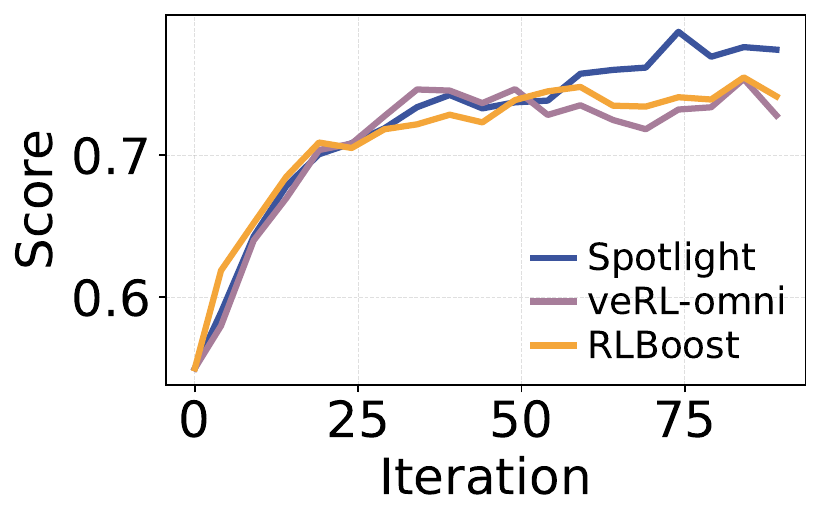}
        \caption{Geneval, $512\times512$.}
        \label{fig:exp_geneval_512}
    \end{subfigure}
    \begin{subfigure}[b]{0.49\linewidth}
        \includegraphics[width=\linewidth]{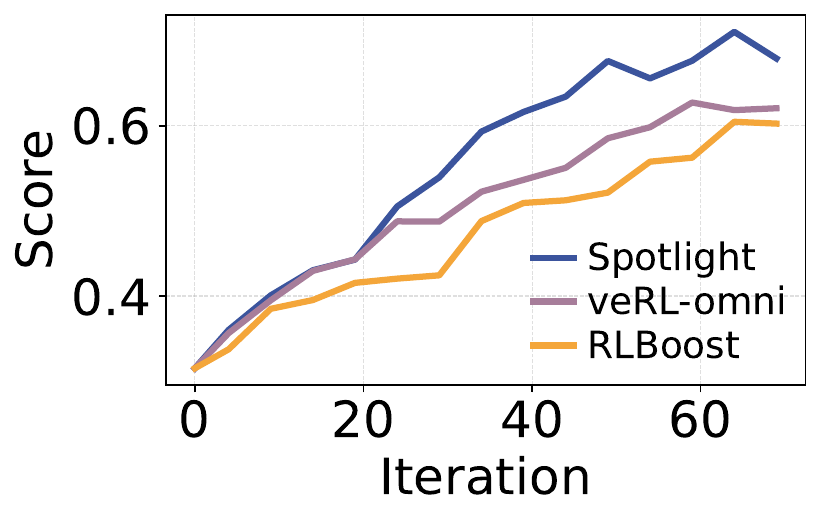}
        \caption{DeepSeek-OCR, $1280\times1280$.}
        \label{fig:exp_ocr_1280}
    \end{subfigure}
    \begin{subfigure}[b]{0.49\linewidth}
        \includegraphics[width=\linewidth]{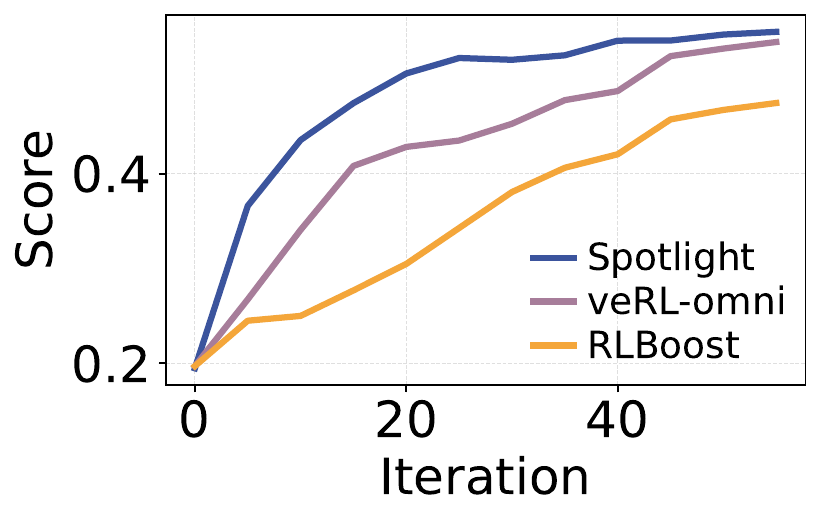}
        \caption{Geneval, $1280\times1280$.}
        \label{fig:exp_geneval_1280}
    \end{subfigure}
    \caption{\textbf{[Dynamic Exploration]:} Validation score vs.\ number of training iterations for \SystemName{}, \Bs, and \Bes on \ocr{} and \geneval{} at $512\times512$ and $1280\times1280$ resolutions. 
    \SystemName{} converges to higher scores across all datasets, using dynamic exploration.
    }
    \label{fig:exp}
\end{figure}

\begin{figure}
    \centering
    \includegraphics[width=\linewidth]{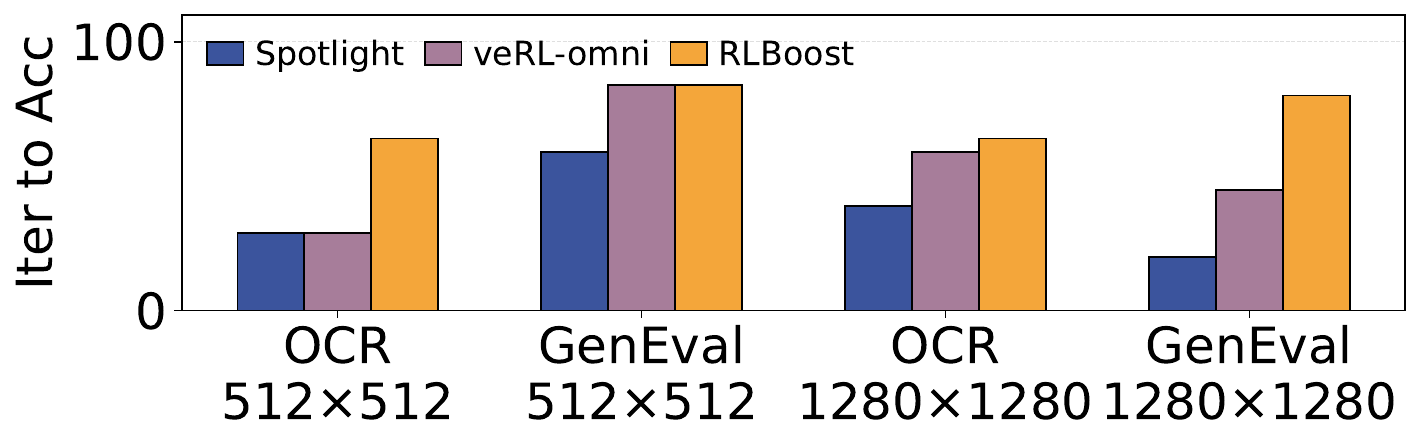}
    \caption{\textbf{[Dynamic Exploration]:} Training iterations required to reach target validation scores. \SystemName{} converges faster than all baselines by dynamic exploration.}
    \label{fig:iter_to_acc}
\end{figure}

Figure~\ref{fig:exp} compares training convergence across all four dataset-resolution configurations. \B{} and \Be{} exhibit identical convergence curves as they differ only in resource provisioning, hence we select \Bs{} and \Bes{} to represent the baseline convergence behavior.

As shown in Fig.~\ref{fig:iter_to_acc}, \SystemName{} reaches the target validation scores using $29$, $59$, $39$, and $20$ iterations for the four configurations respectively, faster than or comparable to all baselines. Compared with \Bs, \SystemName{} converges $1.42$--$4.00\times$ faster because dynamic exploration filters random seeds to retain only those yielding high-contrast rewards, which accelerates convergence. Compared with \Bes, which also performs seed exploration under a fixed configuration, \SystemName{} converges $1.00$--$2.25\times$ faster. This improvement stems from dynamic exploration continuously selecting the seed configuration that maximizes reward variance at each iteration.

\PHM{Overhead Analysis.} As discussed in \S\ref{sec:explore_on_spot}, although \SystemName{} tries to limit exploration within the training time window, this goal cannot always be met for two reasons. First, spot GPU availability changes unpredictably during exploration, altering the effective throughput mid-flight. Second, the discrete action space of exploration hyperparameters does not always contain a configuration that fits exactly within the window. When exploration does not finish in time, the remaining requests are continued by all rollout workers after the model update. We quantify this overhead below.
\begin{figure}
    \centering
    \includegraphics[width=\linewidth]{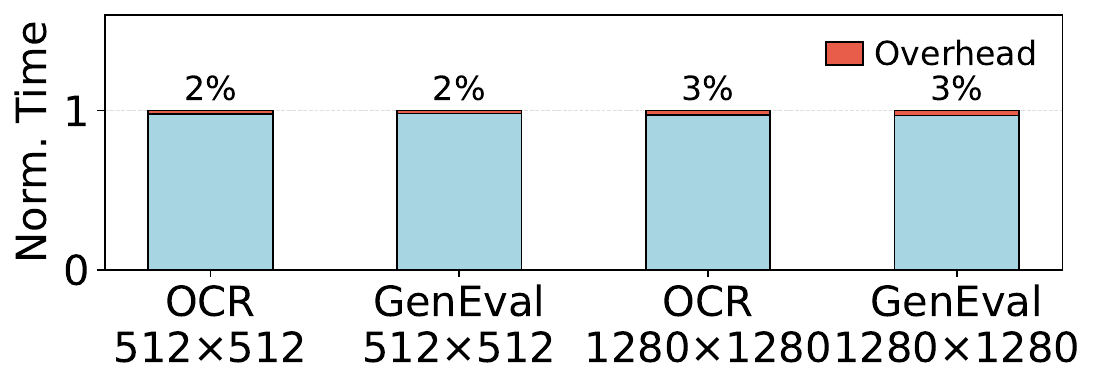}
    \caption{\textbf{[Dynamic Exploration]:} Exploration overhead normalized to average per-iteration time. The overhead remains low across all datasets and resolutions.}
    \label{fig:explore_overhead}
\end{figure}
As shown in Fig.~\ref{fig:explore_overhead}, the exploration overhead remains between $2$--$3\%$ of the average step time across all configurations. This low overhead is due to accurate estimation of spot GPU throughput during the training window and the fact that all rollout workers collectively drain any remaining exploration requests immediately after the model update.

\subsection{Elastic Sequence Parallelism: Robust Throughput}
\label{sec:elastic_sp}
\begin{figure}
    \centering
    \begin{subfigure}[b]{0.49\linewidth}
        \includegraphics[width=\linewidth]{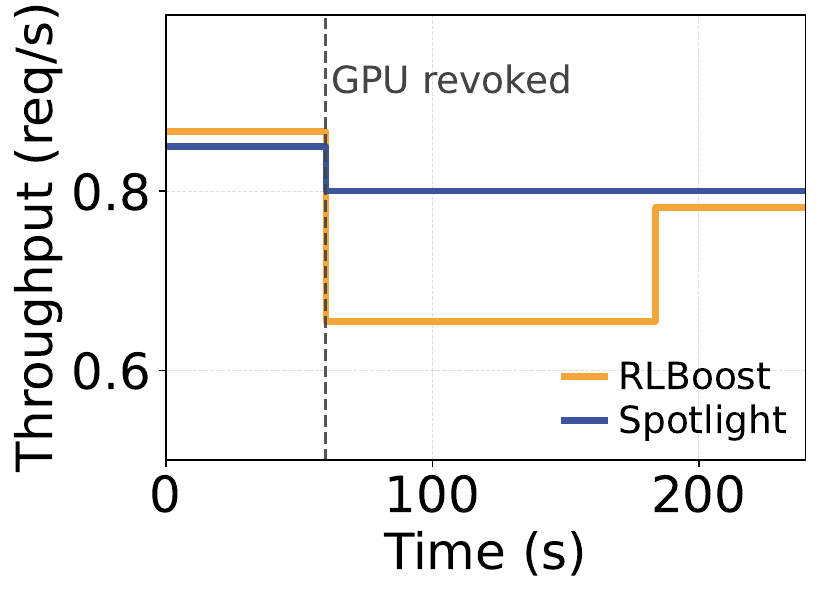}
        \caption{Spot GPU revocation.}
        \label{fig:dsp_minus}
    \end{subfigure}
    \begin{subfigure}[b]{0.49\linewidth}
        \includegraphics[width=\linewidth]{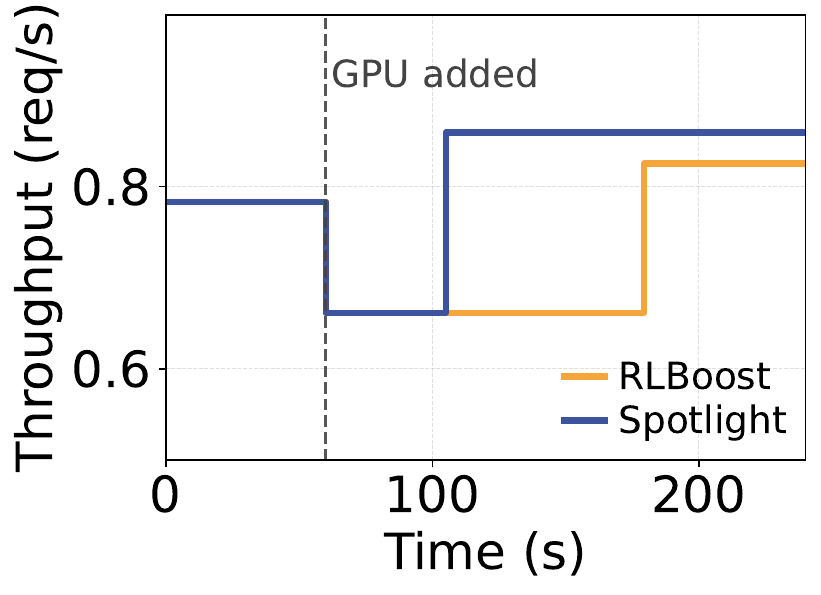}
        \caption{Spot GPU addition.}
        \label{fig:dsp_plus}
    \end{subfigure}
    \caption{\textbf{[Elastic SP]:} 
    Rollout throughput of \SystemName{} and \Bs{} in the presence of revoking (a) and adding (b) one spot GPU.
    }
    \label{fig:elastic_sp}
\end{figure}

To study the effect of elastic sequence parallelism, we zoom into two 240-second windows directly taken from our end-to-end run in \S\ref{sec:e2e}, each containing exactly one spot-capacity change, and compare the rollout throughput of \SystemName{} and \Bs{} side by side. Fig.~\ref{fig:elastic_sp}(a) covers the GPU revocation event at $[831, 1071]$\,s, where the total number of spot GPUs drops from 8 to 7; Fig.~\ref{fig:elastic_sp}(b) covers the GPU addition event at $[1223, 1463]$\,s, where the total number of spot GPUs grows from 7 to 8. We omit \Bes{} from this comparison because its per-step time is too different from that of \Bs{} and \SystemName{}, so its events cannot be aligned with either window for a side-by-side reading.

Upon the GPU revocation event (Fig.~\ref{fig:dsp_minus}), \SystemName{} loses only the affected GPU worker and, with the persistent scheduler still resident, reuses the surviving GPU as an SP=1 worker that offloads parameters to host memory, so the SP transition completes almost instantaneously and rollout throughput drops only marginally; \Bs{} cannot change SP and instead has to terminate the entire SP=2 inference engine on the node, reinitializing a fresh SP=1 engine, which takes about $120$\,s, during which its throughput on that node falls to zero and only catches up with \SystemName{} around $t{\approx}180$\,s into the window. Similarly, on the GPU arrival event (Fig.~\ref{fig:dsp_plus}), \SystemName{} boots one additional GPU worker on the recovering node and lets it copy weights from the existing co-located peer over NVLink, joining the SP group in about $45$\,s; \Bs{} again pays a full engine restart of about $120$\,s before it can absorb the recovered GPU.

In both scenarios, \SystemName{} shows substantially higher throughput robustness than \Bs. Elastic sequence parallelism therefore lets \SystemName{} track spot capacity changes nearly in real time, whereas a naive engine-restart design absorbs them only after a multi-minute delay.

\subsection{Sensitivity to Spot Preemption Frequency}
\label{sec:preemption}
\begin{figure}
    \centering
    \includegraphics[width=\linewidth]{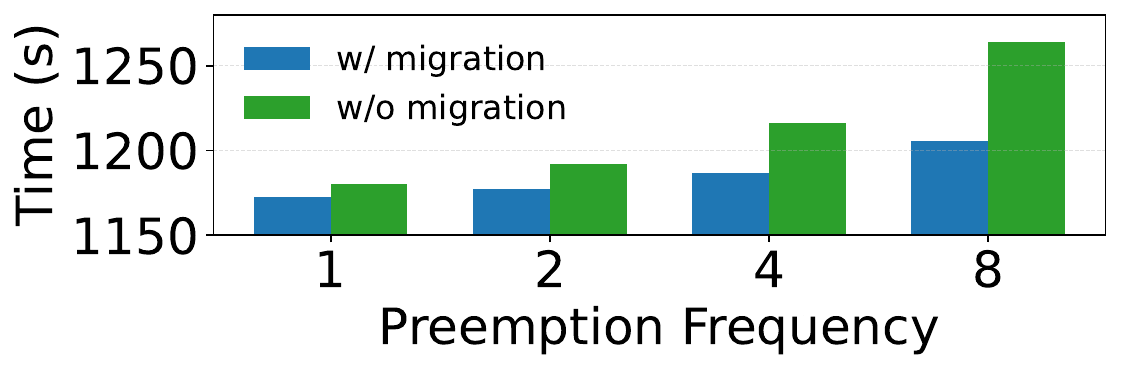}
    \caption{\textbf{[Adaptiveness]:} Iteration duration of \SystemName{} with and without live migration as the number of preemption-recovery cycles per training iteration increases.
    }
    \label{fig:preemption}
\end{figure}
To study the adaptiveness of \SystemName{} to spot preemption, we conduct experiments with varying preemption frequencies. We create a synthetic spot trace where preemption events occur within the time window of one training iteration, each reducing the spot GPU count from 8 to 4 and recovering to 8 after 5\,s. We sweep the preemption frequency from 1 to 8 events per iteration. We evaluate two configurations of \SystemName{}: with live migration enabled and disabled, at image resolution $1280\times1280$.
Fig.~\ref{fig:preemption} shows the iteration duration and overhead under both configurations. When live migration is disabled, overhead comes from recomputing the preempted requests. When live migration is enabled, overhead comes from committing and restoring intermediate tensors to the tensor store on the reserved node. Live
migration allows us to reduce the iteration time by up to $4.6\%$
in the presence of frequent spot GPU revocation, providing
proportionally higher throughput. Note that rollout time increases with preemption frequency due to lower throughput on fragmented GPUs running at SP=1.

\subsection{Ablation Study}
\label{sec:ablation}
\begin{figure}
    \centering
    \includegraphics[width=\linewidth]{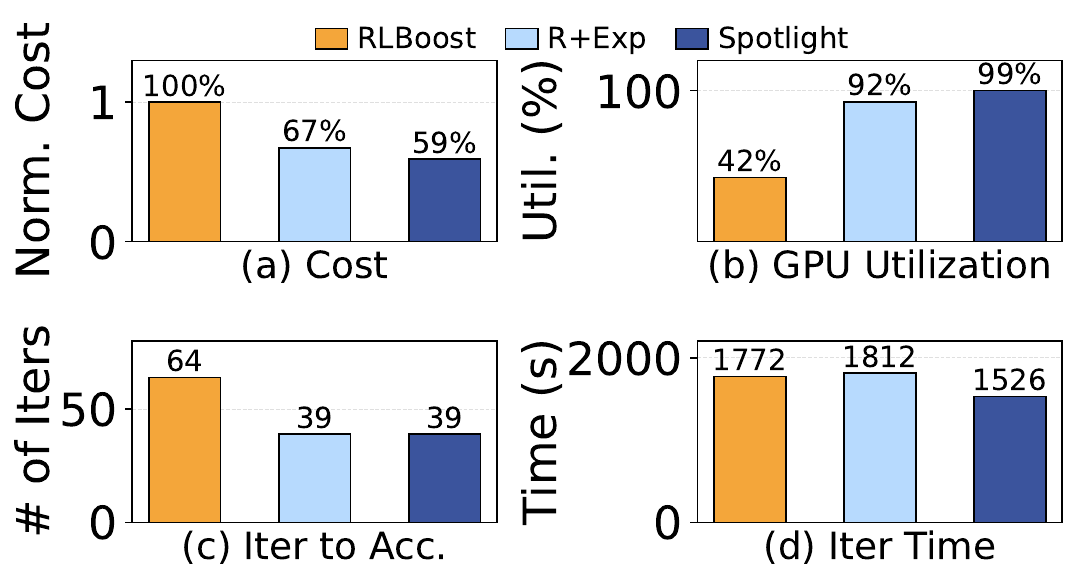}
    \caption{\textbf{[Ablation]:} Ablation study on \ocr{} $1280\times1280$ comparing \SystemName{}, \Bs+Exp (adds dynamic exploration to \Bs), and \Bs. \SystemName{} shows higher GPU utilization and lower iteration number, iteration duration, and cost.
    }
    \label{fig:ablation}
\end{figure}
We conduct an ablation study on \ocr{} at $1280\times1280$ to isolate the contribution of each component. We compare three configurations: \SystemName{} (full system), \Bs+Exp (adds dynamic exploration to \Bs), and \Bs{} (no exploration, no elastic SP).


Dynamic exploration significantly improves training convergence: both \SystemName{} and \Bs+Exp reach the target score with $39\%$ fewer iterations than \Bs{}. However, exploration introduces only $2.3\%$ per-iteration overhead, as discussed in \S\ref{sec:dynamic_explore}. With elastic sequence parallelism, \SystemName{} compensates for this overhead through faster SP reconfiguration, raising spot GPU utilization from $93\%$ to $99\%$ and reducing average iteration time by $15.7\%$ relative to \Bs+Exp. The remaining ${\sim}1\%$ idle time comes from booting on the fresh nodes and GPU worker initialization when increasing the SP degree (\S\ref{sec:elastic_sp}). The combined effect lowers total cost by $1.7\times$.

\subsection{Scalability Study}
\label{sec:scalability}
\begin{figure}
    \centering
    \begin{subfigure}[b]{0.49\linewidth}
        \includegraphics[width=\linewidth]{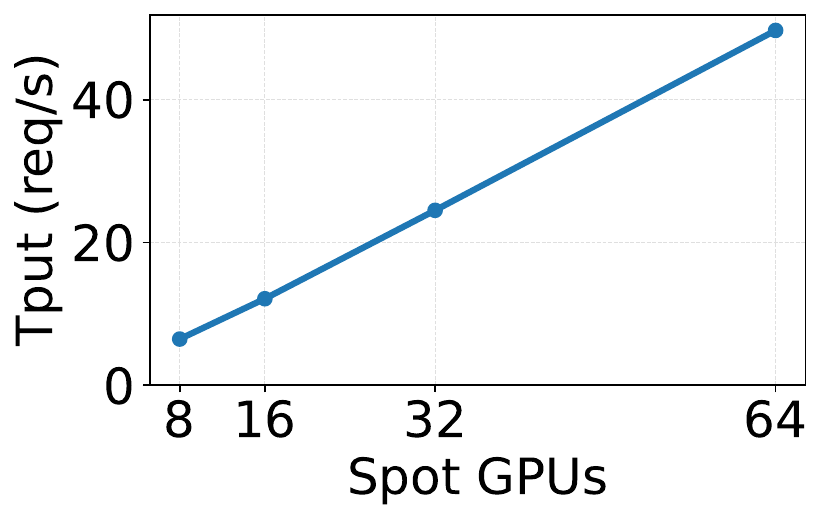}
        \caption{Throughput.}
        \label{fig:scalability_throughput}
    \end{subfigure}
    \begin{subfigure}[b]{0.49\linewidth}
        \includegraphics[width=\linewidth]{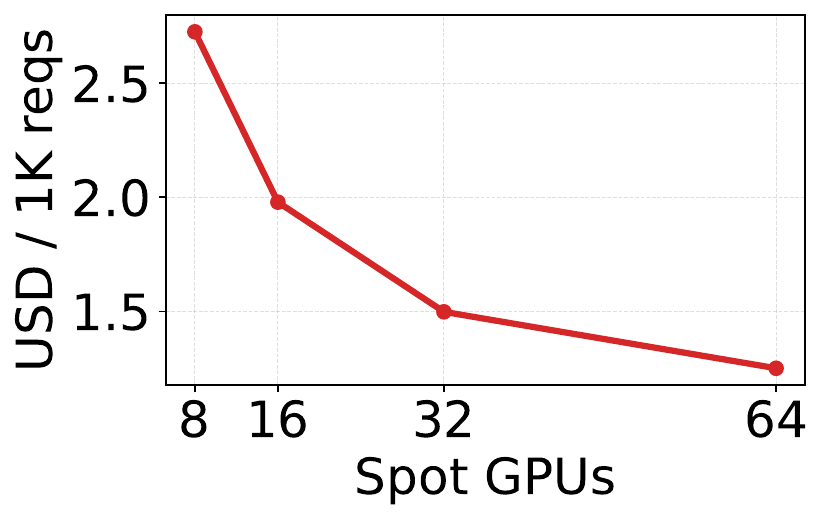}
        \caption{Cost per request.}
        \label{fig:scalability_usd_per_req}
    \end{subfigure}
    \caption{\textbf{[Scalability]:} Throughput and per-request cost as the number of spot GPUs scales.
    }
    \label{fig:scalability}
\end{figure}

We evaluate \SystemName{}'s scalability by fixing four reserved GPUs and varying the number of H100 spot GPUs from 8 to 64 GPUs. To measure peak throughput, we impose no cap on the number of sequences during the exploration phase. We report the effective throughput, including both the rollout and exploration phases, and the cost per request in these two phases. Figure~\ref{fig:scalability} shows that throughput scales from 7 to
50\,req/s when increasing spot GPUs from 8 to 64, achieving a
7.7$\times$ improvement for an 8$\times$ increase in spot capacity.
This near-linear scaling arises because additional spot GPUs increase
the number of exploration requests that can be completed within the
training time window. Meanwhile, cost drops from \$2.72 to \$1.25 per thousand requests, a 2.18$\times$ reduction, because the fixed reserved GPU cost is amortized over more requests and the additional spot GPUs perform effective exploration during the training window.

\subsection{Sensitivity to the \SystemName{} Parameters}
\label{sec:sensitivity}


Two parameters are critical to \SystemName{}'s dynamic exploration (\S\ref{sec:explore_on_spot}): the maximum number of sequences per prompt and the minimum number of denoising steps per sequence. We sweep these two parameters on \ocr{} at $512\times512$ resolution over 20 training iterations.
\begin{figure}
    \centering
    \begin{subfigure}[b]{0.49\linewidth}        \includegraphics[width=\linewidth]{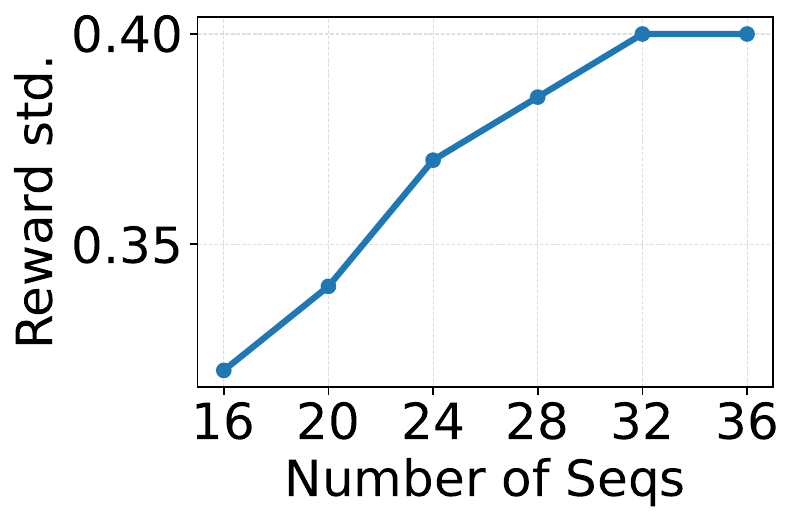}
        \caption{Number of sequences.}
        \label{fig:sensitivity_seqs_max}
    \end{subfigure}
    \begin{subfigure}[b]{0.49\linewidth}
        \includegraphics[width=\linewidth]{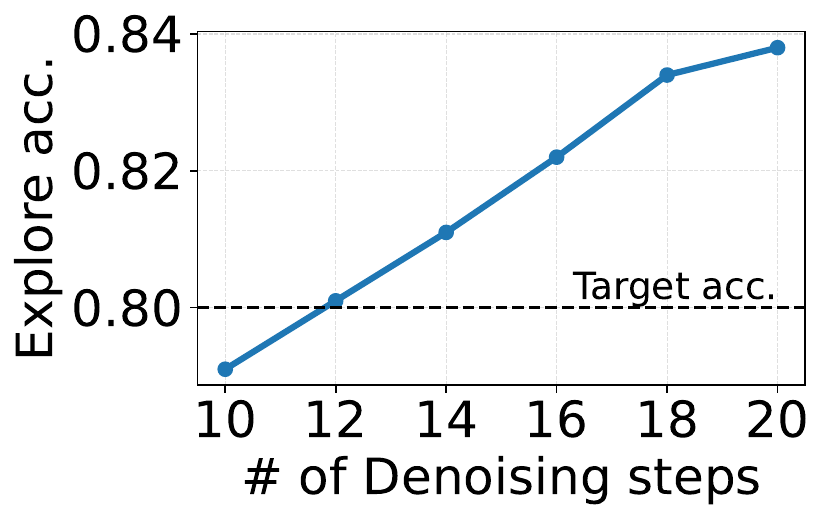}
        \caption{Number of denoising steps.}
        \label{fig:sensitivity_steps_min}
    \end{subfigure}
    \caption{\textbf{[Sensitivity]:} Effect of maximum number of sequences on reward standard deviation, and effect of minimum denoising steps on exploration accuracy. 
    }
\end{figure}
These two parameters together determine the maximum action space of \SystemName{}'s bandit search problem. The maximum number of sequences sets the upper bound on training convergence, as more sequences per prompt yield higher reward variance. We quantify this effect by sweeping the number of sequences during exploration and measuring the per-iteration reward standard deviation. As shown in Fig.~\ref{fig:sensitivity_seqs_max}, reward standard deviation increases with the number of sequences and saturates at $32$. We therefore set the maximum number of sequences to 32.

The minimum number of denoising steps determines the cheapest action in the action space. We aim to use as few denoising steps as possible while preserving exploration accuracy. Following the exploration configuration described in \S\ref{sec:explore_on_spot}, we sweep the number of denoising steps by setting different \tcache{} thresholds and measure exploration accuracy as the rank correlation between exploration rewards and full-rollout rewards. As shown in Fig.~\ref{fig:sensitivity_steps_min}, when the target accuracy is set to 0.8, the minimum number of denoising steps that meets this threshold is 12. We therefore set the minimum denoising steps to 12.
\section{Related Work}

\PHM{Efficient and Cost-Aware RL Systems.}
Existing RL post-training systems primarily improve training efficiency by accelerating rollout generation, optimizing distributed execution, or improving resource utilization. OpenRLHF~\cite{openrlhf}, AReaL~\cite{areal}, veRL~\cite{sheng2024hybridflow}, and ROLL~\cite{roll} provide general execution frameworks that coordinate rollout, reward, and training workers. Subsequent systems optimize complementary parts of the pipeline, including phase composition~\cite{zhong2024rlhfuseefficientrlhftraining,RollMux}, asynchronous execution~\cite{StreamRL,Laminar}, long-tail rollout scheduling~\cite{rollpacker}, and speculative rollout acceleration~\cite{RhymeRL,tlt-adaptive-drafter,seer-sd}.
From a cost perspective, RLHFless~\cite{rlhfless} uses serverless scaling to match dynamic demand in the RLHF pipeline, and RLBoost~\cite{rlboost} offloads LLM rollout to preemptible GPUs. Overall, the systems above mainly target LLM RL, where autoregressive rollouts create long-tail latency and load imbalance. However, in DiT RL, each rollout is compute-intensive and can be accelerated with additional GPUs. \SystemName{} targets this opportunity by using low-cost spot GPUs to accelerate DiT RL training.

\PHM{DiT RL and Seed Exploration.}
Prior works~\cite{flowgrpo,dancegrpo,nft} show that applying GRPO~\cite{shao2024deepseekmath} to diffusion models can improve reward-driven visual generation. Since different random seeds can yield substantially different rewards for the same prompt, seed exploration has become an important technique for improving sample efficiency. DanceGRPO~\cite{dancegrpo} selects high-contrast samples from a larger candidate pool, TreeGRPO~\cite{treegrpo} and BranchGRPO~\cite{branchgrpo} explore structured denoising branches, and Sol-RL~\cite{sol-rl} uses cheap exploratory rollouts to identify high-contrast seeds before full-quality rollout. These works improve the algorithmic efficiency of exploration, but exploration still runs before on-policy rollout, leaving it on the critical path and increasing iteration time. \SystemName{} instead offloads stale-weight exploration to low-cost spot GPUs during training, accelerating convergence and reducing the overall cost.

\PHM{Systems on Spot Instances.}
Spot instances provide low-cost GPU resources but can be preempted, and prior systems have explored using them across training and serving workloads. Bamboo~\cite{thorpe2023bamboo} and Parcae~\cite{parcae} make DNN training on spot instances resilient to preemptions, while SpotServe~\cite{spotserve} and SkyServe~\cite{skyserve} use spot instances for low-cost model serving. \SystemName{} tailors the usage of spot GPUs to the specifics of the DiT RL post-training workloads to overlap exploration and training.

\section{Conclusion}
This paper presents \SystemName{}, a cost-efficient system that harvests spot GPUs for DiT RL post-training. We identify two key insights: seed exploration can tolerate stale model weights, and sequence parallelism reconfiguration can reuse existing state. Building on these insights, \SystemName{} introduces dynamic exploration, which offloads seed exploration to low-cost spot GPUs during the training phase using stale weights. \SystemName{} further employs elastic sequence parallelism and preemption-aware request scheduling to improve spot GPU utilization. Extensive experiments demonstrate that \SystemName{} achieves higher training convergence while delivering significant cost savings over state-of-the-art baselines.

\bibliographystyle{plainnat}
{\small
\bibliography{references}
}

\clearpage
\appendix

\begin{figure}[ht]
    \centering
    \begin{subfigure}[b]{0.49\linewidth}
        \includegraphics[width=\linewidth]{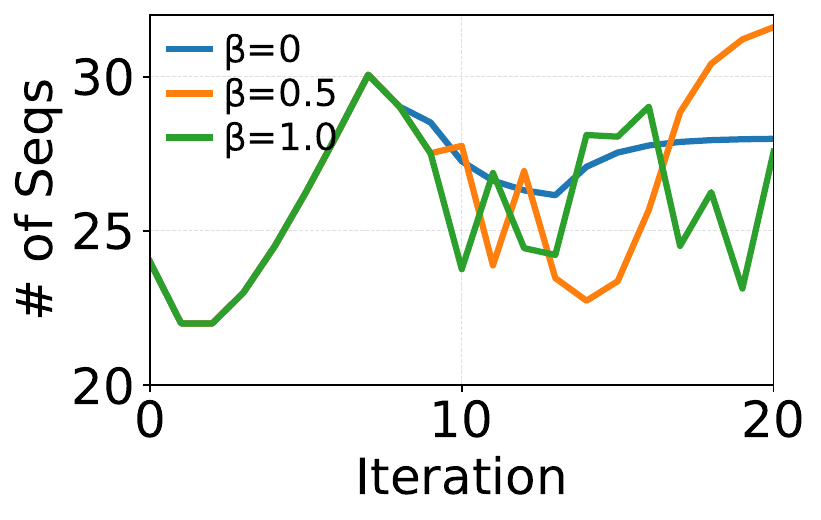}
        \caption{Number of sequences.}
        \label{fig:sensitivity_seqs}
    \end{subfigure}
    \begin{subfigure}[b]{0.49\linewidth}
        \includegraphics[width=\linewidth]{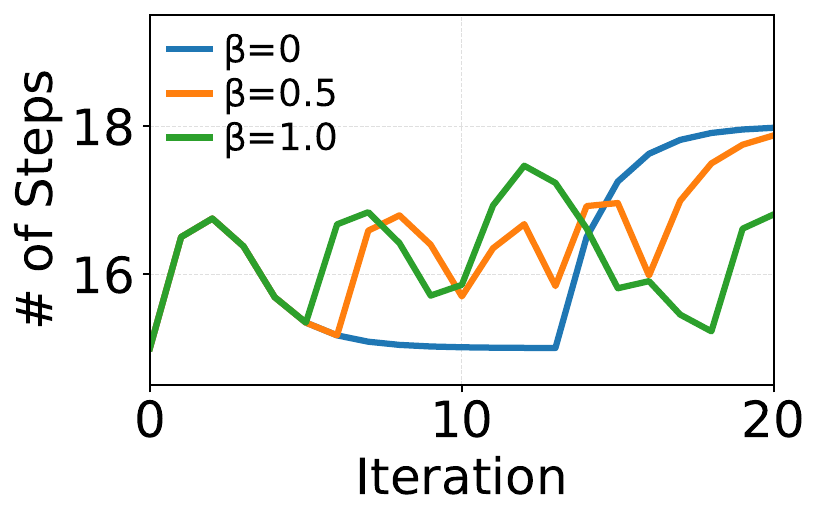}
        \caption{Number of denoising steps.}
        \label{fig:sensitivity_steps}
    \end{subfigure}
    \caption{\textbf{[Hyper Parameters]:} Selected number of sequences and denoising steps over training iterations under different $\beta$ values. Curves are smoothed for visualization. $\beta=0.5$ stabilizes in around 20 steps.}
\end{figure}

\section{Cloud Spot Pricing}
\label{app:spot_pricing}

Following RLBoost's cloud-cost appendix~\cite{rlboost}, we use concrete public-cloud machine prices. Table~\ref{tab:spot_pricing} reports a pricing snapshot queried in June 2026 for H100 instances on AWS, Google Cloud, and Azure~\cite{aws_price_list_api,aws_spot_pricing,gcp_spot_pricing,azure_retail_prices_api}.

The standard entries give on-demand H100 prices, and the spot entries give the corresponding spot H100 prices used in our cost comparison. We normalize every row to per-GPU hourly cost. The average standard H100 price is $(55.04/8+88.49/8+98.32/8)/3=\$10.08$ per GPU-hour. Averaging the four spot entries gives $(14.24/8+2.47+39.81/8+18.17/8)/4=\$2.87$ per GPU-hour, which is $3.5\times$ lower than the standard per-GPU price.

\begin{table}[t]
\centering
\caption{Public-cloud H100 machine prices for cost analysis, following the
RLBoost appendix style. Prices were queried on June 9, 2026.}
\label{tab:spot_pricing}
\small
\resizebox{\linewidth}{!}{%
\begin{tabular}{llccc}
\toprule
Provider & Machine type & \#H100 & Provision & Cost/hour \\
\midrule
\multirow{3}{*}{AWS} & p5.48xlarge & 8 & Standard & \$55.04 \\
 & p5.48xlarge & 8 & Spot & \$14.24 \\
 & p5.4xlarge & 1 & Spot & \$2.47 \\
\midrule
\multirow{2}{*}{GCP} & a3-highgpu-8g & 8 & Standard & \$88.49 \\
 & a3-highgpu-8g & 8 & Spot & \$39.81 \\
\midrule
\multirow{2}{*}{Azure} & ND96isr\_H100\_v5 & 8 & Standard & \$98.32 \\
 & ND96isr\_H100\_v5 & 8 & Spot & \$18.17 \\
\bottomrule
\end{tabular}
}
\end{table}

\section{Hyper Parameters}

\subsection{Convergence Evaluation Hyperparameters}
\label{app:convergence_hyperparams}

Table~\ref{tab:convergence_hyperparams} reports the hyperparameters used for the convergence experiments in \S\ref{sec:dynamic_explore}. We use the same number of sequences per prompt across all configurations and tune the learning rate, weight decay, denoising steps, and SDE window for each dataset-resolution pair.

\begin{table}[t]
\centering
\caption{Hyperparameters used in the convergence experiments.}
\label{tab:convergence_hyperparams}
\small
\resizebox{\linewidth}{!}{%
\begin{tabular}{lccccc}
\toprule
Configuration & Learning rate & Weight decay & \#Seqs & Denoising steps & SDE window \\
\midrule
\ocr{} $512{\times}512$ & $3{\times}10^{-4}$ & $1{\times}10^{-5}$ & 16 & 20 & $[0,15]$ \\
\geneval{} $512{\times}512$ & $1.5{\times}10^{-5}$ & $1{\times}10^{-5}$ & 16 & 10 & $[0,5]$ \\
\ocr{} $1280{\times}1280$ & $3{\times}10^{-5}$ & $5{\times}10^{-5}$ & 16 & 20 & $[0,15]$ \\
\geneval{} $1280{\times}1280$ & $5{\times}10^{-6}$ & 0 & 16 & 10 & $[0,5]$ \\
\bottomrule
\end{tabular}
}
\end{table}

\subsection{Bandit Exploration Coefficient $\beta$}
\label{sec:beta}
The exploration coefficient $\beta$ controls the trade-off between exploitation of well-performing configurations and exploration of under-sampled ones in the UCB-based action selection (\S\ref{sec:explore_on_spot}). We sweep three values: $\beta \in \{0, 0.5, 1.0\}$. As shown in Fig.~\ref{fig:sensitivity_seqs} and Fig.~\ref{fig:sensitivity_steps}, $\beta{=}0$ converges almost immediately within the first 5 iterations, locking into a single configuration too early and failing to explore the broader hyperparameter space. In contrast, $\beta{=}1.0$ continues to oscillate through iteration 20 without settling on a stable configuration. With $\beta{=}0.5$, both the number of sequences and the number of denoising steps converge to stable values around iteration 20, balancing sufficient exploration with timely convergence. We therefore use $\beta{=}0.5$ for all experiments.

\end{document}